\documentclass[preprint]{aastex}  

\slugcomment{submitted to the {\it Astrophysical Journal}}
\shorttitle{Cluster Star Formation and Type Ia Supernovae}
\shortauthors{Loewenstein}

\begin{document}
\title{On Iron Enrichment, Star Formation, and Type Ia Supernovae in
Galaxy Clusters}
\author{Michael Loewenstein\altaffilmark{1}} \affil{Laboratory for
High Energy Astrophysics, NASA/GSFC, Code 662, Greenbelt, MD 20771}
\altaffiltext{1}{Also with the University of Maryland Department of
Astronomy} \email{loew@milkyway.gsfc.nasa.gov}


\begin{abstract}
The nature of star formation and Type Ia supernovae (SNIa) in galaxies
in the field and in rich galaxy clusters are contrasted by juxtaposing
the build-up of heavy metals in the universe inferred from observed
star formation and supernovae rate histories with data on the
evolution of Fe abundances in the intracluster medium (ICM). Models
for the chemical evolution of Fe in these environments are
constructed, subject to observational constraints, for this
purpose. While models with a mean delay for SNIa of 3 Gyr and standard
initial mass function (IMF) are consistent with observations in the
field, cluster Fe enrichment immediately tracks a rapid, top-heavy
phase of star formation -- although transport of Fe into the ICM may
be more prolonged and star formation likely continues to redshifts
$<1$. The source of this prompt enrichment is Type II supernovae
(SNII) yielding $\ge 0.1~{\rm M}_{\odot}$ per explosion (if the SNIa
rate normalization is scaled down from its value in the field
according to the relative number of candidate progenitor stars in the
$3-8~{\rm M}_{\odot}$ range) and/or SNIa explosions with short delay
times associated with the rapid star formation mode. Star formation is
$>3$ times more efficient in rich clusters than in the field,
mitigating the overcooling problem in numerical cluster
simulations. Both the fraction of baryons cycled through stars, and
the fraction of the total present-day stellar mass in the form of
stellar remnants, are substantially greater in clusters than in the
field.
\end{abstract}



\section{Introduction}

Rich clusters of galaxies provide a uniquely amenable setting for the
study of the complex processes and consequences of galaxy formation
and evolution. As the largest ($\gtrsim 3\ 10^{14}~{\rm M}_{\odot}$)
virialized objects in the universe, with the deepest potential wells,
they retain all of their processed and unprocessed baryonic matter in
the form of stars and (predominantly) $\sim 3-10$ keV plasma that are
readily studied at infrared/optical/ultraviolet and X-ray wavelengths,
respectively.

Rich clusters are sufficiently vast to be considered representative
volumes of the universe and are often assumed to be composed of dark
matter, stars, and gas in cosmic proportions. However, because
clusters originate as extreme peaks in the random field of initial
cosmological density fluctuations, the evolution of their constituent
galaxies -- and by extension the intergalactic medium that they are
coupled to -- may proceed in a manner that is distinct from their
counterparts in the field. The resulting biases must be understood,
and taken into account, when generalizing from cluster-based
observables in order to draw conclusions about the universe as a
whole. In fact, the morphological mix, luminosity function, and star
formation history of cluster galaxies all display signatures of the
effects of the exceptionally dense environments where they form and
develop \citep{k02,cr05,s05,rps}. Since galactic outflows associated
with the galaxy formation process, in turn, feed back on the cluster
environment by injecting energy and metals into intergalactic plasma,
evidence of the cluster/field dichotomy is implanted in the
intracluster medium (ICM).

The distinctions between the evolution of baryons in clusters and in
the field may be quantified by the juxtaposition of the build-up of
heavy metals in the universe inferred from the evolving star formation
and supernovae rates with the enrichment history of the ICM in rich
galaxy clusters -- the dominant reservoir for metals produced by stars
in the ensemble of cluster galaxies. Because, under intracluster
plasma conditions, abundances of Fe are more easily measured than
those of other elements they are derived to redshifts $>1$ and provide
the strongest current constraints on the evolution of ICM enrichment.

\citet{ml97} made an initial investigation along these lines based on
their analysis of {\it ASCA} spectra of clusters out to $z=0.3$, and
concluded that ICM enrichment was inconsistent with concurrent
estimates of the star formation rate history derived for field
galaxies (see, also, Madau, Pozzetti, \& Dickinson 1998; Lin \& Mohr
2004; Calura \& Matteucci 2004). Re-examination of this issue is
timely given the extension to higher redshift of ICM Fe abundance
measurements made possible with {\it XMM-Newton} \citep{t03,ha04,m04},
and direct estimates of the evolution of supernova rates to comparable
redshift in the field \citep{s04,d04,bt06} and in clusters \citep{gms}
that constrain the level of supernova metal production per unit star
formation.

Star formation with a standard initial mass function (IMF) produces
approximately one core collapse supernova (hereafter, SNII) per
hundred solar masses of stars formed, while recent empirical estimates
for the specific rate of Type Ia supernovae (SNIa) are an order of
magnitude lower. For typical adopted Fe yields of $0.07~{\rm
M}_{\odot}$ for SNII (slightly higher than the empirical estimate of
Elmhamdi, Chugai, \& Danziger 2003) and $ 0.7~{\rm M}_{\odot}$ for
SNIa \citep{i99}, the estimated universal fraction of baryons in stars
of $\sim 0.06$ \citep{fp04} thus implies an Fe enrichment per baryon
of $\sim 0.065/(1-R)$ relative to solar (solar Fe mass fraction
$=0.0013$) that originates in comparable portions from the two classes
of supernovae, where $R$ is the stellar mass loss fraction integrated
over the age of the universe. Since $R\sim 40$\% for a standard IMF
(see below), one expects a mean universal Fe enrichment per baryon to
about one-tenth solar. While this may be consistent with the
low-redshift (non-cluster) IGM (Prochaska 2004; also, see \S 4.1
below), it falls short by a factor of $\sim 4$ in the ICM (see, also,
Pagel 2002). A simple way to account for this shortfall is to invoke a
top-heavy IMF \citep{dfj,a92,eav,mg95,lm96,gm97} to enhance the
formation efficiency of SNII, and perhaps SNIa. Additional variations
(some linked to the IMF) include the following: (1) a higher average
star formation rate (more SNII and SNIa), (2) a higher incidence of
SNIa per star formed, (3) a higher average Fe yield per (Type Ia
and/or Type II) supernova, (4) a significant enrichment by a
pregalactic stellar population (e.g., Loewenstein 2001). The relative
dearth of supernovae in clusters out to the highest redshifts for
which the rate is presently constrained highlights the cluster
abundance paradox, and begs the question of whether one or more of the
standard assumptions (e.g., an IMF that is universal and invariant
over time) break down in the universe {\it in general}.

This paper aims to evaluate the plausibility and consequences of
possible explanations for ICM Fe enrichment, taking into account
constraints based on the characteristics of stellar populations in
cluster galaxies as well as the observed mean evolution of the cluster
SNIa rate and Fe abundance; and, to examine the resulting implications
for our understanding of the star formation history (SFH) in the
universe and its dependence on environment.

New data on the {\it histories} of star formation, supernovae rates,
and elemental abundances motivate an approach that goes beyond
consideration of the baryon and metal inventories to explicitly model
their self-consistent evolution. The framework adopted here for
constructing such models, as presented in \S 2, is kept simple to
restrict the number of possible parameters while allowing for
consideration of an extensive range in each parameter, and to maintain
the transparency of the effects and implications of various
assumptions and scenarios. Boundary conditions, source functions, and
parameters are initially designed as appropriate for the universe as a
whole; but; clear paths for adaptations that might be required for
rich galaxy clusters are provided. Models for the chemical evolution
of Fe in the ICM are presented in \S 3, and the requirements on the
nature of star formation and supernovae in cluster galaxies imposed by
observations are indicated and summarized. \S 4 includes discussion of
the distinction between clusters and the field with respect to their
intergalactic media and galaxy populations, and more detailed analysis
of galactic outflows and Fe enrichment in clusters. Implications for
the nature of SNIa and its environmental dependence, comparisons with
results of similar investigations, and possible directions for future
model enhancements are also presented in this section. Conclusions,
with an emphasis on the dichotomy between clusters and the field, are
summarized in \S 5. I adopt a topologically flat cosmology with Hubble
constant $H_o=70$ km s$^{-1}$, and total and baryonic matter
densities, relative to critical, $\Omega_m=0.3$ and $\Omega_b=0.045$,
respectively (e.g., Fukugita \& Peebles 2004).

\section{Cosmic Chemical Evolution}

\subsection{Basic Equations}

I consider the coupled evolution of the globally averaged densities
and metal abundances of the three following baryonic categories: (1)
{\bf stars} that includes main sequence and evolved stars, substellar
objects, and stellar remnants; (2) {\bf interstellar gas} (ISM)
defined as the fraction of total matter in the gas phase where stars
may form at any time in the history of the universe (regardless of
location); and, (3) {\bf intergalactic gas} (IGM) defined as the
remaining, inert (e.g., non-star-forming) fraction of gas-phase
matter\footnote{Under these definitions, some (small) fraction of the
``IGM'' may actually consist of hot galactic halo gas, and any gas
originally in circumgalactic or intergalactic space that subsequently
accretes onto galaxies and forms stars is considered ``ISM.''}. Stars
and ISM are coupled via star formation and mass return, the ISM and
IGM via galactic winds (and, potentially, galactic infall -- not
explicitly considered here). Type Ia and core collapse supernovae
directly enrich the ISM. The evolution equations \citep{t80,mg95,tgb}
in rest-frame time for the comoving densities of the three cosmic
constituents -- $\rho_{\rm stars}$, $\rho_{\rm ISM}$, $\rho_{\rm IGM}$
-- and for the mass fractions of their ``ith'' element -- ${f^i}_{\rm
stars}$\footnote{Note that ${f^i}_{\rm stars}$ includes a contribution
from metals locked up in remnants that may be significant
\citep{fp04}.}, ${f^i}_{\rm ISM}$, ${f^i}_{\rm IGM}$ -- are as
follows:

\begin{equation}
{{d\rho_{\rm stars}}\over {dt}}=\dot\rho_{\rm SF}-\dot\rho_{\rm MR},
\end{equation}
\begin{equation}
{{d\rho_{\rm ISM}}\over {dt}}=-\dot\rho_{\rm SF}+\dot\rho_{\rm
MR}-\dot\rho_{\rm GW},
\end{equation}
\begin{equation}
{{d\rho_{\rm IGM}}\over {dt}}=\dot\rho_{\rm GW},
\end{equation}
\begin{equation}
{{d{f^i}_{\rm stars}}\over {dt}}={{\dot\rho_{\rm SF}}\over {\rho_{\rm
stars}}} ({f^i}_{\rm ISM}-{f^i}_{\rm stars}),
\end{equation}
\begin{equation}
{{d{f^i}_{\rm ISM}}\over {dt}}={{\dot\rho_{\rm MR}}\over {\rho_{\rm
ISM}}} ({f^i}_{\rm stars}-{f^i}_{\rm ISM}) +{{\dot{\rho^i}_{\rm
SNIa}}\over {\rho_{\rm ISM}}}+{{\dot{\rho^i}_{\rm SNII}}\over
{\rho_{\rm ISM}}},
\end{equation}
and
\begin{equation}
{{d{f^i}_{\rm IGM}}\over {dt}}={{\dot\rho_{\rm GW}}\over {\rho_{\rm
IGM}}} ({f^i}_{\rm ISM}-{f^i}_{\rm IGM}),
\end{equation}
where the source terms ${\dot\rho_{\rm SF}}$, ${\dot\rho_{\rm MR}}$,
and ${\dot\rho_{\rm GW}}$ are the comoving global mass density rates
of star formation, mass return, and galactic wind mass transfer from
the ISM to the IGM, respectively; and, $\dot{\rho^i}_{\rm SNIa}$ and
$\dot{\rho^i}_{\rm SNII}$ are the density injection rates into the ISM
of element ``i'' from Type Ia and core collapse (mostly Type II)
supernovae, respectively. Since the universe is a ``closed box,''
there are five independent variables; $\rho_{\rm stars}+\rho_{\rm
ISM}+\rho_{\rm IGM}=\Omega_b\rho_{\rm crit}$, where $\rho_{\rm
crit}\equiv 3H_o^2/8\pi{\rm G}$ is the present-day critical
density. Also, the total average baryonic abundance is
\begin{equation}
{f^i}_{\rm bar}={{\rho_{\rm stars}{f^i}_{\rm stars}+\rho_{\rm
ISM}{f^i}_{\rm ISM}+\rho_{\rm IGM}{f^i}_{\rm IGM}}\over
{\Omega_b\rho_{\rm crit}}}.
\end{equation}

\subsection{Functions and Parameters}

The goal of this paper is to explore the chemical evolutionary
implications of the increasingly complete and detailed inventory of
matter, and inferred star formation and metal production rates, in the
universe; and, to evaluate their applicability to the environment of
rich galaxy clusters. As such, empirical quantities and constraints
are directly applied to the fullest possible extent, and pegged as
points of departure in modeling ICM enrichment.

\subsubsection{The Stellar Initial Mass Function}

The initial mass function of forming stars (IMF) $\phi(m)\equiv dN/dm$
must be specified to determine the normalization of the observed star
formation rate and total stellar mass, the stellar mass fraction
recycled to the ISM via mass loss from evolved stars, and the number
of supernova explosions per unit mass of star formation.  Following
Kroupa (2001), I adopt a four-part monotonically decreasing piecewise
power-law IMF, extending from $m_l=0.01~{\rm M}_{\odot}$ in the
substellar regime to $m_u=150~{\rm M}_{\odot}$ \citep{wk04,f05}, with
slope ($-dlogN/dlogm$) $\alpha_0=0.3$ below the hydrogen-burning mass
threshold $m_H=0.08~{\rm M}_{\odot}$, slope $\alpha_1=1.3$ between
$m_H$ and $m_B=0.5~{\rm M}_{\odot}$, and slopes $\alpha_2$ between
$m_B$ and $m_D$ and $\alpha_3$ above $m_D$. $\alpha_2=\alpha_3=2.3$ is
adopted as a standard; alternatives are assumed to have a single-slope
above $m_B$ ($\alpha_2=\alpha_3$), or an additional break at $m_D\ge
2m_B$ where the slope changes from $\alpha_2=2.3$ to $\alpha_3$.

\subsubsection{Mass Return}

The mass density rate of material recycled from stars to the ISM is
\begin{equation}
{\dot\rho_{\rm MR}}(t)={\int_{m_{to}}}^{m_u}dm\phi(m)\Delta m(m)
\dot {n_{\rm SF}}(t-t_{\rm ms}(m)).
\end{equation}
The turn-off mass $m_{to}$ is implicitly given by $t_{\rm
ms}(m_{to})=\Delta t$ where $t_{\rm ms}(m)$ is the main sequence
lifetime of a star of mass $m$ \citep{s92} and $\Delta t$ is time
since the onset of star formation. $\Delta m(m)$ is the mass returned
by stars of main sequence mass $m$, and is assumed independent of
time. The star formation rate density $\dot {n_{\rm SF}}\equiv
{\dot\rho_{\rm SF}}/{{\int_{m_l}}^{m_u}}dm\phi(m)m$.

$\Delta m(m)$ is derived using appropriate remnant masses, $m_{\rm
rem}$, in the white dwarf, neutron star, and black hole regimes
\citep{pcv,vg97,fk01,whw} as follows:
\begin{equation}
\Delta m(m)=m-m_{\rm rem}(m);
\end{equation}
\begin{equation}
m_{\rm rem}=0.45+0.119m,\ m\le 8;
\end{equation}
\begin{equation}
m_{\rm rem}=1.4,\ 8\le m\le 21;
\end{equation}
\begin{equation}
m_{\rm rem}=1.4+0.717(m-21),\ 21\le m\le 33; 
\end{equation}
\begin{equation}
m_{\rm rem}=10, m\ge 33,
\end{equation}
 where $m$ is in ${\rm M}_{\odot}$. 

The integrated fraction of mass formed into stars that is returned to
the ISM up to the present epoch ($t_{\rm now}=13.47$ Gyr) is

\begin{equation}
R={{{{\int_{m_{to}(t_{\rm now})}}^{m_u}}dm\phi(m)\Delta m(m)}\over
{{\int_{m_l}}^{m_u}dm\phi(m)m}},
\end{equation}
where $m_{to}(t_{\rm now})\sim 0.9~{\rm M}_{\odot}$ is the present
epoch main sequence turn-off mass. Under the instantaneous recycling
approximation (IRA: $t_{\rm ms}(m)\rightarrow 0$ for all $m$),
${\dot\rho_{\rm MR}}(t)=R{\dot\rho_{\rm SF}}$.  The exact treatment
(equation 8) is adopted for calculating $\dot\rho_{\rm MR}$, since
star-formation histories that decline on short timescales are
considered. However, in deriving equations 4 and 5, the rate of metal
transfer to the ISM via stellar mass loss at time $t$ is estimated as
$\dot\rho_{\rm MR}(t){f^i}_{\rm stars}(t)$ (rather than by evaluating
an integral similar to that in equation 8).

\subsubsection{Star Formation}

Estimates of ${\dot\rho_{\rm SF}}$, corrected for dust, are compiled
and compared in \citet{h04}, and fitted to analytic functions in
\citet{b05} and \citet{s04}. The star formation rate is renormalized
to assure that
\begin{equation}
{\int_{t_{\rm form}}}^{t_{\rm now}}dt{\dot\rho_{\rm SF}}(t)=
{{\rho_{\rm stars}({t_{\rm now}})-\rho_{\rm stars}({t_{\rm
form}})}\over {(1-R)}},
\end{equation}
where $\rho_{\rm stars}({t_{\rm now}})$ is the cosmic density in stars
observed today, and $\rho_{\rm stars}({t_{\rm form}})$ the density of
zero-metallicity (Population III) ``seed'' stars predating the time,
$t_{\rm form}$, when galaxy formation begins and the integration of
the set of equations (1)--(6) is initiated.

\subsubsection{Core Collapse Supernovae}

The mass density injection rate of element ``i'' into the ISM by core
collapse supernova of massive stars (``SNII'') is
\begin{equation}
{{\dot\rho_{\rm SNII}}}^i(t)={\int_{m_1}}^{m_2}dm\phi(m){y_{\rm
SNII}}^i(m) \dot {n_{\rm SF}}(t-t_{\rm ms}(m)),
\end{equation}
where $m_1<m<m_2$ is the range of SNII progenitor masses and ${y_{\rm
SNII}}^i(m)$ is the nucleosynthetic yield of element ``i'' from a
progenitor of mass $m$. Instantaneous enrichment is an adequate
approximation for these short-lived stars, reducing equation (16) to
\begin{equation}
{{\dot\rho_{\rm SNII}}}^i(t)=K_{\rm SNII}{\dot\rho_{\rm SF}}(t)
\langle {y_{\rm SNII}}^i\rangle,
\end{equation}
where $K_{\rm SNII}\equiv
{{\int_{m_1}}^{m_2}}dm\phi(m)/{{\int_{m_l}}^{m_u}}dm\phi(m)m$ is the
number of SNII per unit mass of star formation, and $\langle {y_{\rm
SNII}}^i\rangle\equiv {{\int_{m_1}}^{m_2}}dm\phi(m){y_{\rm
SNII}}^i(m)/{{\int_{m_l}}^{m_u}}dm\phi(m)$ the mean SNII yield of the
``ith'' element.

\subsubsection{Type Ia Supernovae}

In light of the multiplicity of theoretical predictions for the SNIa
rate (Barris \& Tonry 2006, and references therein) I adopt the
semi-empirical formalism of \citet{s04} and \citet{d04}, determining
the mass density injection rate of element ``i'' by SNIa into the ISM
from
\begin{equation}
{{\dot\rho_{\rm SNIa}}}^i(t)=\dot {n_{\rm SNIa}}(t){y_{\rm SNIa}}^i,
\end{equation}
and
\begin{equation}
\dot {n_{\rm SNIa}}(t)=\nu{\int_{t_{\rm form}}}^tdt'{\dot\rho_{\rm
SF}}(t')\Phi(t-t'),
\end{equation}
where $\nu$ is the number of SNIa progenitor systems per unit mass of
star formation, ${y_{\rm SNIa}}^i$ is the SNIa nucleosynthetic yield
of element ``i'' (assumed to be constant), and $\Phi(t_d)$ is the
normalized delay time distribution function (DTDF) parameterized using
the Gaussian distribution found by \citet{s04} and \citet{d04} to
explain the observed evolution of the SNIa rate,
\begin{equation}
\Phi(t_d)=(2\pi\sigma^2)^{-1}e^{-[(t_d-t_c)^2/2\sigma^2]}.
\end{equation}

\subsubsection{Star-Formation-Induced Galactic Wind}

I assume that supernova explosions drive outflow of material from the
ISM to the IGM and that the mass loss rate per unit volume is
proportional to the total supernova rate and the ISM density,
\begin{equation}
{\dot\rho_{\rm GW}}(t)=K_{\rm GW}(\dot {n_{\rm SNIa}}(t)+K_{\rm
SNII}{\dot\rho_{\rm SF}}(t))\rho_{\rm ISM}.
\end{equation}
Metal-rich gas is not preferentially ejected, and SNIa and SNII are
assumed to contribute to driving outflows with equal efficiency. Given
an observationally determined present-day ISM density, $\rho_{\rm
ISM}({t_{\rm now}})$, setting the galactic wind strength, $K_{\rm GW}$
is equivalent to assuming a value for the initial ISM density,
$\rho_{\rm ISM}({t_{\rm form}})$. I calculate this relationship by
integrating equation (2) from $t_{\rm now}$ to $t_{\rm form}$. In the
absence of infall, $\rho_{\rm ISM}({t_{\rm form}})$ must be sufficient
to account for all the star formation since $t_{\rm form}$, $\rho_{\rm
stars}({t_{\rm now}})-\rho_{\rm stars}({t_{\rm form}})$, and is
bounded above by the total baryon density. That is, allowed values of
$K_{\rm GW}$ are those that yield
\begin{equation}
\rho_{\rm ISM,min}<\rho_{\rm ISM}({t_{\rm form}})<\rho_{\rm ISM,max},
\end{equation}
where $\rho_{\rm ISM,min}=\rho_{\rm ISM}({t_{\rm now}})+(\rho_{\rm
stars}({t_{\rm now}})-\rho_{\rm stars}({t_{\rm form}}))$ corresponds
to no wind, and $\rho_{\rm ISM,max}=\Omega_b\rho_{\rm crit}-\rho_{\rm
stars}({t_{\rm form}})$ to a wind with maximum integrated mass outflow
(``maximum wind'').

Although ram-pressure stripping may be important in determining the
abundance gradient, it is generally thought to account for a small
(though significant) fraction of total ICM metal enrichment (e.g.,
Domainko et al. 2006).  However, it may play a role in extending the
epoch of mass transfer from galaxies to the ICM. I consider an
additional galactic outflow term that may (at least in part) be
associated with this mechanism (see \S\S 3.1.1, 4.2).

\subsubsection{Boundary Conditions}
Equations (1), (2), (4), (5), and (6) are integrated from $t_{\rm
form}$ to $t_{\rm now}$, with metal-free initial conditions:
${f^i}_{\rm stars}={f^i}_{\rm ISM}={f^i}_{\rm IGM}=0$. A small seed
stellar density is assumed, $\rho_{\rm stars}({t_{\rm
form}})=\Omega_{\rm III}\rho_{\rm crit}$; $\rho_{\rm ISM}({t_{\rm
form}})$ is a free parameter within the limits of equation (22).

\subsubsection{Standard and Varying Parameters and Assumptions}

Standard boundary conditions include the formation epoch of the first
Population II stars ($t_{\rm form}=4.66\ 10^8$ yr, corresponding to
redshift 10 for the adopted cosmology), the (Population III) stellar
density $\Omega_{\rm III}\rho_{\rm crit}=3\ 10^{-6}\rho_{\rm crit}$
\citep{og96}, and the stellar ($\rho_{\rm stars}({t_{\rm
now}})=0.00267\rho_{\rm crit}$) and ISM ($\rho_{\rm ISM}({t_{\rm
now}})=0.00078\rho_{\rm crit}$) densities at the present epoch
\citep{fp04}.

The IMF parameters are the slope at high mass, $\alpha_3$, and
transition mass, $m_D$, to this slope from $\alpha_2=2.3$. The choice
of IMF, in turn, determines the mass return fraction $R$ (equation 14)
and renormalization factor for the star formation rate (equation 15),
and the number of SNII per unit mass of star formation, $K_{\rm
SNII}$. In the standard model, $\alpha_3=2.3$, $R=0.393$ (the IRA
value is 0.405), and $K_{\rm SNII}=0.0104~{\rm M}_{\odot}^{-1}$ for
$m_1=8~{\rm M}_{\odot}$ and $m_2=m_u=150~{\rm M}_{\odot}$. The
renormalization factors for the \citet{b05} and \citet{s04} star
formation rate parameterizations are 0.94 and 0.62, respectively. I
adopt the \citet{b05} SFH parametrization as standard. Since the
\citet{s04} rate differs most dramatically in shape at low redshift
where a relatively small fraction of the integrated star formation
occurs, our results are insensitive to this choice (the renormalized
functions are compared in Figure 2; see below). Adoption of the
standard $K_{\rm SNII}$ and ${\dot\rho_{\rm SF}}(t)$ provides
consistency with observed SNII rate evolution to redshift 1 (Strolger
et al. 2004, Dahlen et al. 2004; see, also, Figure 3 below). There is
sufficient theoretical uncertainty in the SNII synthesis of Fe that
the IMF-averaged SNII Fe yield, $\langle {y_{\rm SNII}}^{\rm
Fe}\rangle$, is simply left as a parameter with standard value
$0.07~{\rm M}_{\odot}$.

Standard SNIa parameters include mean delay time $t_c=3$ Gyr,
dispersion $\sigma=0.2t_c$, and normalization $\nu=1.5\ 10^{-3}$ SNIa
${{\rm M}_{\odot}}^{-1}$ (see, also, Gal-Yam \& Maoz 2004, Greggio
2005) that provide the best fit to the observed SNIa rate evolution
for the \citet{b05} star formation rate parametrization (these
slightly differ from their values in Dahlen et al. 2004 due to the
different adopted star formation rate). I focus on Fe in this paper
(thus justifying the neglect of non-explosive production in equation
5); a SNIa Fe yield ${y_{\rm SNIa}}^{\rm Fe}=0.7~{\rm M}_{\odot}$ is
adopted as standard.

For the above standard parameters and functions, one may vary the
galactic wind factor from $K_{\rm GW}=0$ to ${K_{\rm GW}}^{\rm max}$,
where $\Omega_b\rho_{\rm crit}{K_{\rm GW}}^{\rm max}=3.05\ 10^3~{{\rm
M}_{\odot}}$ is such that all of the IGM originates in the ISM.

\subsubsection{A Note on Varying the IMF}

The standard model IMF (with slope 2.3 above $0.5~{\rm M}_{\odot}$),
if assumed universal, provides mutual consistency among observations
of the star formation and SNII rates and build-up of stellar mass
\citep{b05,do05,gh05}, and with the luminosity density of the universe
\citep{bg03}. If the slope is significantly flatter, (1) the higher
stellar mass loss return fraction (Figure 1) implies that a higher
star formation rate than is observed would be required to produce the
observed amount of stars\footnote{Neglecting any effects of a
different IMF on the observational estimates of the stellar mass.}
(Figure 2); and, (2) the implied SNII rate would be greater than
observed (Figure 3a). Conversely, a steeper high mass IMF slope (i.e.,
$\alpha_2=2.8$; see Kroupa \& Weidner 2003) significantly
underpredicts the observed SNII rate \citep{wk05} unless the
transition at $m_D$ is pushed to high mass (Figure 3b).

\subsection{Enrichment in the Standard Model}

The chemical evolution of Fe in the standard model (see entries
``1N2.3'' in Tables 1 and 2) provides predictions to be compared with
abundance measurements in stars and interstellar gas in field
galaxies, and in the intergalactic medium. This serves to indicate
possible shortcomings in the model and provide a baseline for
evaluating what manner of extension or variation might be required to
explain Fe abundances in rich galaxy clusters. Integrated over a
Hubble time, the total stellar Fe yield
\footnote{Note that this is the yield per star formed and is not
renormalized to the present stellar mass.} (relative to solar) in the
standard model is 1.3 (58/42\% from SNIa/SNII) corresponding to an
enrichment of 0.13 averaged over all baryons (Table 2). The
model-predicted $z=0$ SNII and SNIa rates (Table 2) are consistent
with observations \citep{cet}.

The distribution of Fe among stellar, ISM, and IGM components as a
function of time depends on the strength and time-dependence of
galactic outflow. The more numerous SNII dominate the galactic wind
term (equation 21), but SNIa contribute the majority of Fe enrichment
-- and do so with a significant lag with respect to the accumulation
of stellar mass and the ejection of SNII-enriched material into the
IGM. This results in a relatively recent ($z<1$) build-up of Fe in the
stars (Figure 4a) and ISM (Figure 4b) and a substantial Fe mass
fraction contained in these components at the present epoch.  As a
result, little more than half of the total Fe production is injected
into the IGM for a ``maximum'' wind (Figure 5). A reapportionment that
boosts the IGM Fe fraction requires extending the outflow duration, a
possible mechanism considered in detail below.

A comparison, focusing on the $z>2$ universe, of the enrichment
buildup in these models with the inventory of metals derived from
observations of damped Ly$\alpha$ systems, the Ly$\alpha$ forest, and
Lyman break galaxies is made in \S 4.1 below.

\section{Modeling Fe Enrichment of the Intracluster Medium}

The formalism of the previous section can be directly adapted to
investigate the chemical evolution of stars, ISM, and IGM (ICM) in
clusters -- assuming that they are closed boxes -- by simply rescaling
the physical densities by the mean cluster baryon overdensity factor
(i.e., the total overdensity multiplied by a cluster baryon ``bias''
factor -- see, e.g., Allen et al. 2004), $\delta\sim 10^2$. Clearly,
the standard model must be adjusted in application to clusters as it
predicts an IGM Fe abundance $>5$ times smaller than measured in the
ICM.

There is strong evidence that star formation ensued more quickly in
clusters with respect to the field, as one might expect in regions of
highest initial overdensity \citep{s05,rps}. However, this in itself
does not result in enhanced enrichment if the integrated star
formation and IMF are unaltered. Estimates for the rich cluster mass
fraction in stars vary, but are typically twice the estimated
universal value of $\sim 0.06$ (see \S 4.4). Nevertheless, the Fe
enrichment shortfall persists if the yield per star formed is that of
the standard model \citep{p02,n05a}. Moreover, there are claims of
still higher stellar mass fractions in galaxy groups \citep{pa05,e05};
yet, group IGM Fe abundances are not generally higher than in the ICM
(though groups may not be closed boxes). Therefore, as elaborated on
below, a top-heavy IMF may be required to explain the observed 0.4
solar ICM Fe abundance.

While much of the star formation in intermediate and low mass galaxies
occurs after the peak in the universal SFH at $z\approx 1$, the most
massive (elliptical) galaxies were assembled and their star formation
completed at earlier epochs
\citep{sa04,b05,f04,mc04,h05,t05,c05,ji05,pap06}. Since the fraction
of stars formed in elliptical galaxies is higher in clusters and star
formation is expected to be accelerated in denser environments
\citep{k04,t05}, I consider an enhancement in star formation, referred
to as the ``rapid mode'', with an exponential time-dependence,
$\dot\rho_{\rm XSF}\propto e^{-t/(\tau_Xt_{\rm now})}$. These stars
may reside in protogalactic fragments, or even in isolation in
intracluster space \citep{lm04,zgz}.  Star formation in these extended
models is characterized by the dimensionless parameter $\tau_X$, the
normalizations of the ``normal'' (i.e., with the same time-dependence
as the average field star formation rate; see Figure 2) and rapid star
formation modes (expressed, in Table 1, as the present-epoch stellar
baryon fractions: $f_{\rm stars}$ and $f_{\rm Xstars}$, respectively),
and the IMF of each mode (i.e., $m_D$, $\alpha_3$, $m_{DX}$,
$\alpha_{3X}$; in practice, when considering a nonzero exponential
star formation contribution, $\alpha_3=2.3$ is often adopted). For the
models discussed in the next section $\tau_X=0.1$ is chosen.

\subsection{Accounting for Fe, and its Evolution, in Clusters}

\subsubsection{General Considerations}

If the ratio of intergalactic gas to stars in clusters is the same as
the universal average, $\sim 16$, severe difficulties, as described in
\S 2, emerge in constructing models that reproduce observed cluster Fe
abundances. This is illustrated in Figure 6, where the solid curve
plots the relationship between the ratio of the Fe abundance in
galaxies (i.e., a mass-weighted average of stars and ISM) to that in
the ICM versus the mass fraction of Fe in the ICM, assuming the
universal IGM fraction of 0.924. A conservative upper limit to the
ratio of galactic-to-intergalactic Fe abundances of 5 requires that
$>70$\% of the Fe reside in the ICM. However, as indicated above for
the standard model, if galactic winds are driven by star formation
while enrichment is more prolonged due to the delay in SNIa
explosions, then more than half of the Fe is locked up in galactic
stars. This results in a galaxy/ICM Fe abundance ratio {\bf $>12$}
(Figure 6). This problem is exacerbated if (as expected) star
formation is accelerated in clusters relative to the field, but is
mollified if star formation is more efficient. A
galactic-to-intergalactic Fe abundance ratio of 5 implies an ICM Fe
mass fraction $\sim 55$\% ($\sim 45$\%) for a stellar mass fraction
that is twice (three times) the universal average (see dotted and
dashed curves if Figure 6).

To transport sufficient Fe from galaxies to the ICM requires an
outflow that is extended in time if the standard treatment of SNIa
rates and yields, as previously defined, is adopted. I thus modify
equation (21) by adding one of the two following terms: (1)
${\dot\rho_{\rm GW}}(t)\rightarrow {\dot\rho_{\rm GW}}(t)+\rho_{\rm
ISM}/t_{\rm wind}$ (additional exponential outflow term), or (2)
${\dot\rho_{\rm GW}}(t)\rightarrow {\dot\rho_{\rm GW}}(t)+\rho_{\rm
ISM,min}/\Delta t_{\rm wind}$ (additional constant outflow term),
where $\rho_{\rm ISM,min}$ is defined following equation (22) and
$t_{\rm wind}$ and $\Delta t_{\rm wind}$ are additional parameters
(that modify the relationship between $K_{\rm GW}$ and $\rho_{\rm
ISM}({t_{\rm form}})$).

\subsubsection{Results of Representative Models}
The equations of \S 2 are solved for an extensive multi-dimensional
grid of SFHs ($f_{\rm stars}$, $m_D$, $\alpha_3$, $f_{\rm Xstars}$,
$m_{DX}$, $\alpha_{3X}$) and outflow ($t_{\rm wind}$ or $\Delta t_{\rm
wind}$, $K_{\rm GW}$ or -- equivalently -- $\rho_{\rm ISM}({t_{\rm
form}})$) parameters. Conservatively, further consideration is
restricted to models that produce ICM Fe mass fractions of $40-80$\%.
Integrated supernovae Fe yields per baryon of $\sim 0.5-1$ relative to
solar are thus implied, given the 0.4 solar ICM Fe abundance. The
outflow strength and evolution must be adjusted to release sufficient
Fe into the ICM from galaxies, and affects how quickly the high ICM Fe
abundance accumulates.

Of course, there is a large parameter degeneracy among models that
predict 0.4 solar ICM Fe abundance at $z=0$. These are constructed and
evaluated in the context of recent data on cluster Fe abundance and
SNIa rate {\it evolution}, not necessarily to narrowly determine these
parameters, but to reveal the general distinctions between these
models and those constructed to explain standard universal chemical
evolution. We will see that none of the models are fully satisfactory,
thus motivating exploration of additional variations.

Details of an illustrative cross section of models that are examined
in more detail are displayed in Table 1. These are labeled by (1) the
present-day total stellar mass fraction relative to the universal
value (there are equal contributions from normal and rapid modes in all
hybrid SFH models), and (2) whether the SFH follows the normal (N) or
rapid (R) mode functional form, or a hybrid (H) of the two. They are
further denoted by (3) the high mass IMF slope ($\alpha_3$ or
$\alpha_{3X}$), where the rapid mode slope is used for the hybrid SFH
models (except for model 1H0.8Wc where $\alpha_3=1.8$, the normal mode
slope is 2.3 for all H-models; Table 1). The break mass, $m_D$ or
$m_{DX}$, is usually $2~{\rm M}_{\odot}$ and introduces no additional
ambiguity otherwise. Finally (4), the designation ``Wx'' (``Wc'')
denotes whether an extended exponential (constant) outflow term is
included, with the normalized timescale displayed in Table 1. The
choice of outflow form and timescale, as well as the parameter
$\rho_{\rm ISM}({t_{\rm form}})$ (expressed in Table 1 as the initial
baryon mass fraction in the ISM, $f_{\rm ISM}({t_{\rm form}}$)),
likewise introduce no ambiguities -- {\it i.e.} an optimal value is
selected for each SFH/IMF combination. The models appearing above the
demarcation line in Tables 1 and 2 follow the standard SNIa and SNII
treatments explained in \S 2, and have $t_{\rm form}$ corresponding to
$z=10$. This subset demonstrates the diversity of predicted model
outcomes and level of consistency with cluster data, and how these
vary with changes in the input parameters.

The present-day characteristics of these models are displayed in Table
2, and the evolution of the ICM Fe abundance back to $z=1.5$ for these
models is illustrated and compared to observations in Figures
7ab. Since infall is not considered, ISM abundances are significantly
higher than stellar abundances (see Figure 4; also, Figure 8b below;
and, Pipino et al. 2005). Since one might expect higher metallicity
ISM to form stars more readily and equalize these, I define the
``galactic'' abundance, ${\langle{\rm Fe}\rangle}_{\rm gal}$, as the
mass-weighted average of the ISM and galaxy Fe abundances (see Table
2); ${\langle{\rm Fe}\rangle}_{\rm gal}$ serves as an upper limit to
the stellar Fe abundance.

The following general statements can be made. (1) As discussed above,
a model with the form of the universal SFH, the universal fraction of
baryons in stars, and standard transformation from star formation rate
to supernova enrichment, falls many factors short of providing
sufficient Fe (e.g., Model 1N2.3; solid curve in Figure 7a). (2) As
illustrated by Model 2H1.05 (dotted curve in Figure 7b), basic models
where the outflow strictly traces the supernova rate generally predict
fractions of Fe locked up in stars (and ISM) that are too high to
yield consistency with present-day ICM Fe abundances (for a reasonable
galactic abundance, see discussion above). (3) Flattening the IMF
whilst retaining the standard model SFH form (e.g., Model 1N1.05Wc;
dotted curve in Figure 7a) produces the observed $z=0$ ICM Fe
abundance if an extended outflow is assumed; however, the buildup of
Fe is too recent to be consistent with X-ray spectroscopy of clusters
over redshifts 0--1 \citep{t03,ha04,m04}. This is also the case in
other models with insufficient massive star formation at high
redshift, including Model 1H0.8Wc where the universal value for the
baryon fraction in stars originates equally in normal and rapid star
formation modes (short-dashed curve in Figure 7a).

The observed magnitude and mild evolution of the ICM Fe abundance is
fairly well reproduced in a variety of models with a top-heavy IMF,
prompt star formation, and extended outflow. These include Models
1R1.05Wc and 2R1.55Wx where all star formation is in rapid mode
(long-dashed and dot-long-dashed curves in Figure 7a, respectively),
and in a number of hybrid (``2H'') models that include both star
formation modes.

The ``2H'' series of models (see Tables 1 and 2, and Figure 7b) are
those characterized by SFHs producing twice the universal average of
the present-day baryon fraction in stars (see \S 4.4), with half
originating in normal star formation mode with a standard IMF and half
in rapid mode with a steeper high mass IMF, that yield $z=0$ Fe
abundances $\sim 0.4$ solar. These provide fairly good explanations of
the observed ICM Fe abundances over $z=0-1$, provided that the IMF is
sufficiently top heavy and the outflow is of sufficient magnitude and
duration (Models 2H1.05Wx, 2H1.3Wx, 2H1.55Wx). Earlier ICM enrichment
in models with exponential winds, compared to that in models with
constant winds, results in a flatter decline in Fe abundance with
redshift and a better match to the observed ICM Fe evolution. The
effect of varying the supernova wind strength over the full physically
allowed range is illustrated in Figures 8ab; Fe abundances span a full
range $\Delta{\langle{\rm Fe}\rangle}_{\rm ICM}\sim 0.1$ at each
redshift.

The SFHs for Models 1N2.3, 1N1.05Wc, 2H1.3Wx, and 2R1.55Wx are shown
in Figure 9 (the latter two are representative of the best fits to the
observed Fe abundance evolution from Figures 7ab, the former two are
included for comparative purposes). SFHs in dual-mode (hybrid) models
follow the (scaled) field SFH out to $z=1$ since a standard IMF is
assumed for the normal mode. The star formation rates in the models
with a rapid component, of course, greatly exceed the field value at
high redshift; their SFHs resemble those inferred in massive galaxies
\citep{he04,yhg}. Similar enhancements are realized in the Type II and
Type Ia supernova rates (Figures 10ab), although the SNIa explosion
delay leads to a turnover in the rate at high redshift. The solid
curves (Model 1N2.3) match the rates in the field, while the errorbars
are cluster rates measured by \citet{gms}. The latter are converted
from units of ``SNU'' (supernovae per 100 years per $10^{10}$ solar
$B$-band luminosities) assuming that clusters have twice the overall
stellar mass-to-light ratio and twice the stellar fraction as in the
field, and show surprising consistency with the field rate
evolution. Large intermediate redshift SNIa rates in clusters are
predicted in these models where the abundance and evolution of Fe in
the ICM is explained via SNII and SNIa from a top-heavy IMF
enhancement in high redshift star formation, and where the SNIa
delay-time characteristics and relative normalization are the same as
in the field. These rates are marginally inconsistent with the
empirical cluster estimates, a characteristic shared by all models of
the type discussed up to this point that successfully explain the ICM
Fe abundance and its evolution.

Additional models with empirically motivated variations in formation
epoch and various supernova input parameters are described below the
demarcation line in Tables 1 and 2, and discussed in the following
three subsections.

\subsubsection{Reducing the Mean Supernovae Delay Time}

In order to construct models that better agree with the cluster SNIa
data, I consider a reduction, from 3 to 0.5 Gyr, in the mean SNIa
delay interval \citep{mg04} that may be applied to all SNIa (Model
2H1.05WxSt1 in Tables 1 and 2). Mannucci, della Valle, \& Panagia
(2006) argue for a bimodal population of SNIa that includes one
subclass that is associated with active star formation and
characterized by delay times that are insignificant relative to the
duration of the star formation epoch, and another subclass associated
with quiescent secular evolution that is characterized by a broad
distribution extending to long delay times. I also, therefore,
consider models where the reduction in mean delay time is exclusively
applied to SNIa originating in rapid star formation mode (Model
2R1.55WxSt -- with only rapid mode star formation, Model 2H1.05WxSt2).
The results are shown in Table 2 and Figures 11ab. Agreement with
observations of the ICM Fe abundance evolution is significantly
improved, and the $z<1$ SNIa rate histories -- that now more closely
trace the SFHs -- in the hybrid SFH models are in better accord with
the observational constraints.

\subsubsection{Reducing the Formation Redshift}

Since the heretofore considered rapid star formation models appear to
underpredict the present-day cluster SNIa rate, I consider a sequence
of rapid SFH models where the initiation of star formation is moved
forward from $z(t_{\rm form})=10$ to $z(t_{\rm form})=3$. The outcomes
of three such models (2R1.55WxSz, 2R1.55WxStpz, 2R1.55WxStz; Table 1),
distinguished by mean SNIa delay times ($t_{cX}$) that are,
respectively, 3, 1.5, and 0.5 Gyr are summarized in Table 2 and
Figures 12ab. The SNIa rate evolution in these models, optimized to
match the observed cluster Fe abundance evolution (Figure 12a,) is
compared to that in models with earlier formation epoch in Figure
12b. Delaying the onset of star formation improves the match to
cluster SNIa rate observations for $t_{cX}=0.5$ Gyr, but worsens it
for $t_{cX}=3$ Gyr.

\subsubsection{Naturally Reducing the SNIa Rate}

I also construct models with ``naturally'' reduced rapid star
formation mode specific SNIa rate -- {\it i.e.} the scaling is
calculated assuming that a universal fraction of $3-8~{\rm M}_{\odot}$
stars forms SNIa progenitor binaries. I consider models characterized
by $m_{DX}=m_D=2~{\rm M}_{\odot}$, $\alpha_3=2.3$, $t_c=3$ Gyr,
$z(t_{\rm form})=10$ or 3, $t_{cX}=3$ or 0.5 Gyr, and $\langle {y_{\rm
SNII}}^{\rm Fe}\rangle=0.07$ or $0.10~{\rm M}_{\odot}$; and, focus
here on an illustrative subset of ``2H'' and ``2R'' models (i.e.,
where 12\% of present-day cluster baryons are in stars formed either
in equal proportions in normal and rapid mode, or exclusively in rapid
mode) with $\alpha_{3X}=1.05$ and 1.55, respectively. The
corresponding rapid mode SNIa rate reduction factors are 0.26 and
0.66.

In order for these models with reduced early SNIa enrichment to match
the $z>0.3$ ICM Fe abundance evolution as well as, e.g., the models
presented in Figures 11a and 12a, an increase in mean SNII Fe yield to
(at least) $\langle {y_{\rm SNII}}^{\rm Fe}\rangle=0.10~{\rm
M}_{\odot}$ is required (Figure 13a). Because the early ICM Fe
enrichment is now dominated by SNII, the ICM Fe evolution is less
sensitive to the choice of $t_{cX}$ (for $\langle {y_{\rm SNII}}^{\rm
Fe}\rangle=0.07~{\rm M}_{\odot}$, the fall-off of ICM Fe with redshift
in 2H models is steeper if $t_{cX}=3$ Gyr than if $t_{cX}=0.5$ Gyr).
Since the differences in the SNIa rate history emerge beyond the
redshift where observed constraints are available (Figure 13b), the
SNIa delay time in clusters is unconstrained by observations if the
rapid star formation in cluster galaxies was characterized by a flat
IMF and high SNII Fe yields. Note that with most of the Fe enrichment
due to SNII in the rapid star formation mode, an extended outflow
phase is no longer required -- provided that the initial ISM fraction,
$f_{\rm ISM}({t_{\rm form}})$, is large (e.g., model
2R1.55Stnz). Inclusion of a prolonged outflow period reduces the
required value of $f_{\rm ISM}({t_{\rm form}})$.
 
\subsubsection{Model Result Overview}

A simple model for the chemical evolution of Fe in the universe may be
constructed using parameters drawn from the observed star formation,
and Type II and Type Ia supernova, rates in the field population of
galaxies that are mutually consistent for a standard IMF. For Fe
yields of $0.07~{\rm M}_{\odot}$ per SNII and $0.7~{\rm M}_{\odot}$
per SNIa, the baryons in the universe are enriched to an average Fe
abundance of 0.13 solar. The universal ratio of
galactic-to-intergalactic baryons constrains the fraction of Fe locked
up in stars and cold gas such that stellar Fe abundances are within a
factor of two of solar with the precise value dependent on the
magnitude and timing of galactic outflows. The prolonged epoch of star
formation coupled with the SNIa delay shifts much of the Fe enrichment
to relatively recent ($z<1$) epochs, as indicated by recent
observations \citep{la06}.

If one naively assumes that rich galaxy clusters are pure
representative samples of the universe, this global chemical evolution
model directly scales to clusters and Fe abundances in the ICM are
badly underestimated. However, in order to better understand the
differences in the nature of star formation in cluster and field
galaxies, it is instructive to adopt the field model as a point of
departure for investigating the variations that are required to
reproduce observed $z\approx0$ Fe abundances of $\sim 0.4$ solar and
mild ($z=0-1$) ICM Fe abundance evolution.

Successful variations must include shifts to {\it both} higher
redshift in the peak of the star formation rate, and to flatter high
mass slopes in the IMF for star formation at high redshift -- implying
that conversion of gas into stars is, on average, more efficient in
rich clusters than in the field.

Overall Fe enrichment originates in roughly comparable fractions from
SNIa and SNII for models where the relationship between SNIa and star
formation is the same in clusters and the field. Due to the long SNIa
delay, outflow of Fe from galaxies extends beyond the epoch of rapid
star formation in order to be incorporated into the ICM; however, if
the outflow is too prolonged, ICM enrichment is too recent and in
conflict with the mild observed Fe evolution (see discussion in \S 4.2
below). The implied SNIa rates at $z\sim 1$ are somewhat at odds with
observational limits for clusters, although the conflict is slightly
weaker than inferred by \citet{mg04} (see discussions in \S\S 4.3 and
4.6 below). Unfortunately, there is as yet no data to provide a more
definitive test by offering a comparison with the very high $z=1-2$
cluster SNIa rates predicted by these models (see Figure 7b; Maoz \&
Gal-Yam 2004).

Additional models with SNIa rate histories that differ from those
above, and that generally improve the match to the $z<1$ evolution in
ICM Fe inferred from X-ray data, are constructed by (1) shortening the
mean delay time in the kernel that transforms the star formation into
the SNIa rate, (2) decreasing the initial star formation redshift in
models where all stars form in rapid mode, or (3) reducing the number
of SNIa explosions per unit mass of stars formed according to the
number of intermediate mass ($3-8~{\rm M}_{\odot}$) stars whilst
increasing the average SNII Fe yield to $0.1~{\rm M}_{\odot}$. The
SNIa evolution in models of these type, with both rapid and normal
star formation modes, most closely reproduce the observed $z<1$
supernova rates. Observations of rapid build-up of Fe, and relatively
low $z<1$ star formation and SNIa rates, in clusters are most readily
explained in models where heavy element enrichment traces star
formation.

\section{Discussion, Implications, Predictions}

\subsection{The IGM Versus the ICM}

It is valuable to compare, in more detail, observed constraints on the
chemical evolution of the universe outside of clusters with the
standard model of \S2, i.e. a model with star formation history,
stellar mass buildup, and Type II and Type Ia supernova rate histories
as observed in the field, and with standard IMF and SNII and SNIa Fe
yields. Such an examination yields insights into the chemical
evolution of the universe, and reveals distinctions from that peculiar
to galaxy clusters. Comparisons for the abundance evolution to $z=6$
of the stellar, ISM, and IGM components are shown in Figure 14.
Allowing for an Fe-to-$\alpha$-element ratio as low as one-third, the
IGM abundance evolution in models with wind parameter greater than
that corresponding to $\Omega_{\rm ISM}({t_{\rm form}})>0.0076$ (see
middle curves in Figure 4 and 5) is consistent with the $z=5$
Ly$\alpha$ forest lower limit \citep{s01}, and with $z=2.5-3.5$
Ly$\alpha$ forest measurements \citep{s03,ssr}. This implies that
$>9$\% of the $z=0$ IGM originates in galaxies and that the average
stellar Fe abundance at $z=2.5$ is $\sim 0.05-0.1$ relative to solar.

The model Fe abundance averaged over all baryons is also consistent
with that measured in damped Ly$\alpha$ systems at $z=0.4-1.5$ (argued
by Rao et al. 2005 as dominating the mean cosmic metallicity in this
redshift interval). The model ISM abundances are consistent with the
$z=2.5-3.5$ damped Ly$\alpha$ measurement if $\Omega_{\rm ISM}({t_{\rm
form}})<0.028$.  The standard model (without an altered IMF, a
temporally extended wind, or adjustment of relative normalizations,
delays, or yields for supernovae) is in general agreement with all
$z>2$ non-cluster metallicity observations (see, also, Daigne et
al. 2005).

Note that, at least for Fe, there is no evidence of the missing metal
problem summarized in, e.g., \citet{pe04}: the models that assume the
empirical $z>2.5$ star formation rate are in accord with the
observations. There are two factors at play here. The primary one is
that, although the Fe yield per star averaged over a Hubble time is
indeed $\sim 1.3$ solar as generally assumed, the average yield prior
to $z=2.5$ is only $\sim 0.6$ (see Figure 4) due to the delay in SNIa
element production. The second factor is that the total ISM density
must exceed the value measured in damped Ly$\alpha$ systems,
$\Omega_{\rm DLA}(z=2.5)\sim 0.0015$ \citep{ph04}. The amount of
potentially star-forming gas must be sufficient to account for the
stars formed over $z=0-2.5$: $\Omega_{\rm ISM}(2.5)>0.0027$, taking
mass return into account and assuming a standard IMF. If one also
accounts for the mass loss from the ISM necessary to account for the
metals observed in the IGM, $\Omega_{\rm ISM}(2.5)>0.0052$ is inferred
-- more than three times the damped Ly$\alpha$ value. Thus, the
deficit in the detected amount of metals relative to expectations may
be better characterized as a ``missing gas (ISM)'', rather than a
``missing metal,'' problem in the $z\sim 2.5$ universe (see, also,
Hopkins, Rao, \& Turnshek 2005; Prochaska, Herbert-Fort, and Wolfe
2005).

Conversely, there {\it would} be a missing metals problem if the
variations considered above needed to explain cluster Fe abundances
and its evolution were applied to the field, as these were
specifically introduced to produce a prodigious rapid
enrichment. Models where half the stars in the field were formed from
a flat-IMF, rapid star formation mode with either reduced SNIa delay,
or increased SNII yields and suppressed SNIa rate, overpredict $z=2.5$
Fe abundances outside of clusters by $>10$. Star formation is
fundamentally different in clusters and the field (see \S 4.4, below).

A full consideration of limits on departures of the standard model
consistent with observations outside of clusters is beyond the scope
of this work. Given the above conclusions, it is perhaps surprising
that field and cluster galaxy populations are not more distinct in
terms of apparent ages, mass-to-light ratios, and abundances/abundance
ratios -- especially since spheroids dominate the stellar mass in both
populations \citep{b03}. A general prediction implied by the
successful cluster models presented here is of higher Fe abundances in
cluster galaxies compared to the field (even though a higher
percentage of Fe is locked up in galaxies in the field -- $\sim 60$\%,
compared to $\sim 40$\% in clusters). This effect is mitigated if
metals are preferentially lost in outflows associated with the rapid
star formation. 

\subsection{The Magnitude and Timing of Galactic Outflows in Clusters}

I find that $>25$\% of the ICM must originate in galaxies ({\it i.e}
the ``ISM'') in order for the ICM to be enriched in Fe to 0.4 solar.
The combination of (1) observations of the large amount of
intracluster Fe and its mild evolution since $z=1$, (2) significant
limits on the cluster SNIa rate since $z=1$, and (3) the assumption
that Fe in the ICM originates in galaxies that are known to form most
of their stars at $z>1$ constrains the amount, epoch, and duration of
matter and metal transport from cluster galaxies to the ICM. Thus
observations of the evolution of ICM enrichment provide a unique
diagnostic of the nature of galactic winds associated with galaxy
formation in a dense environment.

Since SNII are more numerous than SNIa at all times for all models
\footnote{Except at very recent epochs in rapid-only star formation
models, when both rates are low.} (see Figures 10ab, noting the
respective scales), they are primarily responsible for the ejection of
material from galaxies. The SNII rate (and, hence, SNII Fe enrichment)
is proportional to the star formation rate that steeply declines with
redshift in cluster galaxies. Because the mass loss rate as
implemented in equation (21) is proportional {\it both} to the SNII
rate {\it and} to the mass in interstellar gas that decreases with
time as it is consumed by star formation and also lost in galactic
winds, it falls even more quickly than the star formation rate -- and,
more quickly than the interstellar medium is enriched. Therefore a
prompt phase of only modestly enriched mass injection into the ICM
ensues; and, an additional, more prolonged, outflow (see \S 3.1.1)
must be invoked to transport the necessary Fe into the ICM -- even if
the SNIa delay time is reduced to 0.5 Gyr. The sole exceptions to this
requirement are models where SNII fully dominate the ICM Fe enrichment
that {\it may} be assumed to undergo prompt and very strong (implying
very high initial ISM baryon fractions) outflows, although scenarios
with longer duration and milder outflows (lower initial ISM baryon
fractions) are also feasible (see \S 3.1.5). However, in models where
the galactic outflow era is overly prolonged, the resulting delay in
ICM enrichment leads to an underprediction of the observed ICM Fe
abundances at $z>0.3$.  

Such an additional extended phase of mass loss may be identified with
ram-pressure stripping of galaxies rather than galactic winds, since
simulations indicate that this is the dominant ICM enrichment
mechanism at $z<1$ \citep{s05,d06}. In the models presented here with
extended outflow, $\sim 25-40$\% of the enrichment is due to this
phase, compared to $\sim 10-25$\% in the simulations.  Other possible
avenues of prolonging the enrichment timescale include efficient winds
from a galaxy subpopulation (perhaps late-type or dwarf galaxies)
distinct from that responsible for the prompt outflow phase, or the
suppression at early epochs in the conversion efficiency of supernovae
energy to the kinetic energy of outflow (e.g., if it were more
efficiently radiated as a result of a denser protogalactic
environment). Some, but not all, these mechanisms are particular to
the rich cluster environment.

\subsection{Implications for Supernova Physics and Cosmology}

As shown in, e.g., Figure 10b the specific cluster SNIa rate is
similar to that in the field out to $z=1$ -- the maximum redshift for
which observational constraints are available in either
environment. The inference, drawn above, that cluster star formation
is significantly more efficient then implies -- assuming universal
SNIa Fe yields -- that SNIa must occur less frequently per unit mass
of star formed, and/or predominantly explode at $z>1$.

Figures 15ab revisit selected curves from Figures 10b, 11b, 12b, and
13b, focusing on the $z<1.3$ region, and with the SNIa rates on a
linear, rather than logarithmic, scale. The solid lines show models
with standard SNIa normalizations and DTDFs. These are in marginal
agreement with the observed cluster rate, and predict an order of
magnitude increase just beyond the highest observed redshift bin.

For the hybrid star formation models, decreasing the mean delay time
or the number of SNIa per star formed according to the IMF improves
the match to the observations and results in less extreme behavior
just beyond $z=1$ (Figure 15a), while simultaneously providing models
that more precisely fit the ICM Fe data (assuming, in the
reduced-normalization models, that the SNII Fe yield is $>0.1~{\rm
M}_{\odot}$). In contrast, the SNIa rate in models where star
formation exclusively occurs in rapid star formation mode generally
decline too steeply at low redshift.

There is independent justification for characterizing SNIa explosions
associated with the rapid star formation in cluster galaxies with
relatively short delay times. Physically plausible double degenerate
binary system progenitor models for SNIa with short delay times
\citep{bbr} apparently find realization in regions of active star
formation. Evidence includes the measurement of higher SNIa rates in
blue, relative to red, galaxies \citep{m05}; and, of higher rates in
early-type galaxies when they are radio loud \citep{d05}. This is
consistent with the observation of systematic differences in the
characteristics of individual SNIa in elliptical and spiral galaxies
(Della Valle et al. 2005, Garnavich \& Gallagher 2005). \citet{f02}
also suggested that multiple SNIa types are required based on
gradients in abundance ratios in the ICM. High and persistent Fe
abundances, and Fe-to-$\alpha$ ratios, in the nuclei of galaxies
hosting high redshift quasars also may indicate the presence of
short-delay-time SNIa (Dietrich et al. 2003, Maiolino et al. 2003).
The successful models presented here with reduced SNIa delay times in
the rapid star formation mode fit in well with these observations,
providing a connection between star formation in local galaxies and
the starbursts responsible for the elliptical galaxy stellar
populations that dominate in rich clusters. I have shown that, if the
SNIa per star formed in such models is determined by renormalizing the
field value according to the relative number of $3-8~{\rm M}_{\odot}$
stars, then the SNII Fe yield must exceed $0.1~{\rm M}_{\odot}$
irregardless of the rapid mode mean SNIa delay time. However, one need
not assume that progenitors of the two proposed subclasses of SNIa
explode with the same likelihood \citep{mdp}.

Even for a given progenitor model, the distribution of delay times is
sensitive to the distribution of orbital separations and mass ratios,
and other factors, that depend on the initial mass function and galaxy
age and metallicity in a complex and uncertain manner \citep{g05}.
Moreover, the evidence for a 3 Gyr mean delay for SNIa in the field
has recently been questioned \citep{bt06}. Given these considerations,
in combination with uncertainties in the configuration of SNIa
progenitors and explosion mechanism, one hesitates to suggest that the
possibility that rapid star formation produce SNIa with shorter delay
times casts doubt on the utility of SNIa as standard candles
\citep{yl00}. In any case, since models where SNII dominate Fe
enrichment are not ruled out, the evolution of intracluster Fe does
not constrain the SNIa delay time in a model-independent
way. Additional consideration of abundance ratios, and particularly
their evolution, may be more definitive.

\subsection{The Efficiency and Nature of Star Formation in Cluster
Galaxies Compared to Field Galaxies}

The baryon mass fraction in stars, $f_{\rm b,stars}$ and star-to-gas
ratio, $f_{\rm b,stars}/f_{\rm b,gas}$, in rich galaxy clusters may be
estimated from several recent observational studies -- although
uncertainties persist. Estimates of the total B-band mass-to-light
ratio cluster around $300~{\rm M}_{\odot}/{\rm L}_{{\rm B}\odot}$ with
a spread of $\sim 30$\% \citep{mh02,g02,vym,sp03}, while the stellar
mass-to-light ratio for a composite stellar population dominated by
old stars to the degree appropriate for clusters is $\sim 4.5~{\rm
M}_{\odot}/{\rm L}_{{\rm B}\odot}$ (Marinoni \& Hudson 2002; note,
however, that this assumes a standard IMF). This yields a mass
fraction in stars of $\approx 0.015$ and $f_{\rm b,stars}/f_{\rm
b,gas}=1/8$ (assuming a cluster baryon bias factor of 0.9) -- roughly
twice the universal ratios. Similar ratios are derived from estimates
of the total \citep{k03} and stellar \citep{do04} mass-to-light ratios
where luminosities are measured in the K-band (see, also, Lin, Mohr,
\& Stanford 2003).

The overall implications of these estimates of baryon mass fractions
in stars for galaxy clusters that significantly exceed the universal
value, and their connection to the problem of elemental abundances in
the ICM, may not be fully appreciated. The modeling approach presented
here provides a quantitative perspective on some of the distinctions
between star formation in the field and in clusters that complements
studies of stellar populations. In models that successfully reproduce
observed $z=0-1$ ICM Fe abundances, a significant fraction of the
current mass in cluster galaxy stars forms in a rapid mode of star
formation characterized by a top-heavy IMF. The properties of the
populations of Lyman-break and submillimeter galaxies provide
independent evidence for such an IMF in the environments of rapid
merger-driven star formation at high redshift of the kind that
dominates in cluster galaxy progenitors \citep{bl05}. Because of the
higher return fraction in this component, more than two-thirds of the
integrated star formation occurs in this mode, in accord with the
large fraction of cluster stellar mass residing in early-type systems
\citep{g03,p05}, and the high formation redshift
\citep{bl04,ho04,c04,v04,cm04,t05,l05} and relatively small amount of
continuing star formation \citep{j05,t05,rps} in cluster galaxies. The
inferred mass return fraction is $\sim 70$\%, compared to 40\% for a
standard IMF, implying that star formation would be 2--3 times more
efficient than would be the case for a standard IMF even assuming the
same present-epoch stellar mass fraction. If the baryon fraction in
stars at $z=0$ is indeed twice as high in clusters as in the field,
and stars form in field galaxies with a predominantly standard IMF,
then star formation in clusters is {\it 3 -- 5 times more
efficient}. Likewise, $\sim 25-35$\% of cluster baryons may be
inferred to be cycled through stars (this fraction is $\sim 10$\% for
the standard model). Moreover, although the mass fraction of stars
formed that is currently locked up in stellar remnants is $\sim
13-15$\%, $\sim 40-60$\% of the stellar mass {\it today} is in that,
non-luminous, form (compared to $\sim 20$\% for a standard IMF).

These inferences have implications for constructing and interpreting
population synthesis models of cluster galaxies, and for estimating
their stellar mass-to-light ratios (e.g., Zepf \& Silk 1996), as well
as for evaluating the extent and nature of ``overcooling'' and
feedback in numerical simulations of cluster formation
\citep{b01,knv}.  It is crucial not to neglect mass return in
inventories of the integrated star formation, especially in clusters.

Most, but not all, cluster star formation is required to occur at high
redshift. As discussed in the previous section, models that provide
the best simultaneous fits to the SNIa rate and ICM Fe evolution are
those with hybrid star formation histories, since the SNIa rate
steeply declines at $z<1$ in models where star formation exclusively
originates in rapid mode. This may be taken as an indication that the
finite (if small) low-redshift cluster SNIa rates imply a
non-negligible degree of relatively recent star formation activity in
cluster galaxies. While the Butcher-Omeler effect indicates that
recent epoch star formation is suppressed relative to the field by a
factor of $>10$ in rich cluster cores, star formation persists outside
of the core where late-type galaxies are more common \citep{b06}.  In
any case, since the estimated low-$z$ cluster SNIa rate is only
$3\sigma$ above 0, it is likely that one could construct hybrid-SFH
models with a lower level of recent ($z<0.5$ -- considering the delay
in SNIa enrichment) star formation than in those presented here that
would be equally successful -- particularly if one considers a SNIa
delay time distribution function that falls off less steeply at long
delays than the Gaussian function considered here. In fact, if there
are two subpopulations of SNIa, with DTDFs centered on short ($<<10^9$
yr) and long ($>10^9$ yr) delays, respectively, an exponential
distribution for the long-delay SNIa provides a better fit to the
observed field SNIa rate history \citep{mdp}.

\subsection{Origin and Enrichment History of Intracluster Fe}

In the models that, simultaneously, best match the ICM Fe abundance
and SNIa rate evolution, ICM enrichment traces star formation via
either a combination of SNII and short-time-delay SNIa (2H1.05WxSt1,
2H1.05WxSt2; Figure 11), or domination by SNII (2H1.05WxSn,
2H1.05WxStn; Figure 13). For the former, the fraction of stars in SNIa
progenitors is assumed to be the same as for a standard IMF, even
though the total fraction of stars in the $3-8~{\rm M}_{\odot}$ range
is lower. Under the IRA, the integrated SNIa Fe enrichment per baryon,
relative to solar, is
\begin{equation}
{\langle{\rm Fe}\rangle}_{\rm bar}({\rm
SNIa})=\left({0.097\over{1-R}}\right)\left({{{y_{\rm SNIa}}^{\rm
Fe}}\over{0.7~{\rm M}_{\odot}}}\right)\left({{f_{\rm stars}^{\rm
total}}\over {0.12}}\right)\left({{\nu}\over {1.5\ 10^{-3} {\rm SNIa}\
{{\rm M}_{\odot}}^{-1}}}\right),
\end{equation}
where, ${f_{\rm stars}^{\rm total}}=f_{\rm stars}+f_{\rm Xstars}$ is
the total present-day stellar baryon fraction. The cluster value
inferred from observations is
\begin{equation}
{\langle{\rm Fe}\rangle}_{\rm bar}({\rm OBS})=0.346\left({{f_{\rm
ICM}}\over{0.865}}\right)\left({{\langle{\rm Fe}\rangle}_{\rm
ICM}\over{0.4}}\right)+0.135\left({{1-f_{\rm ICM}}\over{0.135}}\right)
\langle{\rm Fe}\rangle_{\rm gal},
\end{equation}
where $f_{\rm ICM}$ is the cluster baryon fraction contained in the
ICM; $f_{\rm ICM}=0.865$ corresponds to $f_{\rm stars}^{\rm
total}=0.12$. For $\langle{\rm Fe}\rangle_{\rm gal}=1-3$ (the
successful models here predict values at the higher end of this
range), ${\langle{\rm Fe}\rangle}_{\rm bar}({\rm OBS})\approx
0.5-0.75$.

${\langle{\rm Fe}\rangle}_{\rm bar}({\rm SNIa})$ is calculated,
without assuming the IRA, and plotted in Figures 16ab as a function of
rapid star formation IMF slope ($\alpha_{3X}$), assuming
$m_{DX}=m_D=2~{\rm M}_{\odot}$, $\alpha_3=2.3$, $z(t_{\rm form})=10$,
$f_{\rm stars}^{\rm total}=0.12$, and ${y_{\rm SNIa}}^{\rm
Fe}=0.7~{\rm M}_{\odot}$. Evidently, SNIa synthesize a significant
fraction of cluster Fe only in models characterized by a flat IMF, and
with $\nu$ maintained at its value deduced in the field. Moreover, the
SNIa Fe yield must be high. While recent multi-dimensional simulations
\citep{t04,ro06} predict lower yields, phenomenological studies
continue to favor higher values of ${y_{\rm SNIa}}^{\rm Fe}$
\citep{ba06}.

Similarly, the integrated SNII Fe enrichment per baryon, assuming IRA,
is
\begin{equation}
{\langle{\rm Fe}\rangle}_{\rm bar}({\rm
SNII})=\left({0.065\over{1-R}}\right)\left({{{y_{\rm SNII}}^{\rm
Fe}}\over{0.07~{\rm M}_{\odot}}}\right)\left({{f_{\rm stars}^{\rm
total}}\over {0.12}}\right)\left({{K_{\rm SNII}\over {10^{-2}{\rm SNII}}\
{{\rm M}_{\odot}}^{-1}}}\right).
\end{equation}
Figures 16ab also include the computation, without assuming the IRA,
of ${\langle{\rm Fe}\rangle}_{\rm bar}({\rm SNII})$ for the parameters
listed above and ${y_{\rm SNII}}^{\rm Fe}=0.07~{\rm M}_{\odot}$, as
well as the summed SN Fe enrichment per baryon. In addition, the local
IMF slope, and range of ${\langle{\rm Fe}\rangle}_{\rm bar}({\rm
OBS})$ estimated above, are indicated. These figures emphasize the
robustness of the requirement for a flat IMF (unless ${y_{\rm
SNII}}^{\rm Fe}>0.2~{\rm M}_{\odot}$), and illustrate the challenge in
producing sufficient Fe in the hybrid star formation models (that best
match the observed cluster SNIa evolution) if the SNIa progenitor
probability for stars in the $3-8~{\rm M}_{\odot}$ range is
universal. The Fe enrichment per baryon for such models plateaus at
$\sim 0.55$, explaining why ${y_{\rm SNII}}^{\rm Fe}=0.1~{\rm
M}_{\odot}$ (or greater) is favored if $\nu$ is scaled in this manner
(\S3.1.5).

To summarize, the observed frequency of SNIa in clusters (or in the
universe as a whole) is insufficient to account for cluster Fe
enrichment. Therefore, either a separate class of prompt, high Fe
yield SNIa that initiate at high redshift explode with an incidence
per star formed that exceeds what one expects based on the field value
and relative number of $3-8~{\rm M}_{\odot}$ stars, or the shortfall
is made up by SNII. These SNII must primarily originate from rapid
star formation with a flat IMF, and yield an average of $>0.1~{\rm
M}_{\odot}$ of Fe. If this is the case, SNII dominate cluster
enrichment, implying a SNII-like pattern of elemental abundances in
the ICM and in the oldest cluster galaxies. This pattern is fixed at
high redshift, unlike its counterpart in field galaxies where Fe
enrichment is delayed. Enhanced numbers of SNII, and likely gamma-ray
bursts as well, in clusters at $z>1.5$ are then expected.

\subsection{Additional Notes on Similar Studies}

\citet{mg04} reached similar conclusions with regard to the inadequacy
of SNIa, and the likely need for a top-heavy IMF, to account for ICM
Fe.  Their approach was essentially as follows. They calculated the
number of SNIa required to make up the Fe deficit left after
subtracting the SNII contribution -- assuming a Salpeter IMF
($\alpha_0=\alpha_1=\alpha_2=\alpha_3=2.35$) -- from what is observed
at $z=0$. They then distribute those SNIa in redshift according to a
particular formation epoch $t_{\rm form}$ and SNIa delay time
distribution -- assuming that all stars formed in a burst at $t_{\rm
form}$ -- for comparison with the cluster SNIa rate evolution derived
from their observations. In agreement with the present work, they find
that the specific incidence of SNIa progenitors required exceeds the
field value by an order of magnitude.  Assumptions in their analysis
that differ from the present one include (1) $z(t_{\rm form})=2$ or 3,
(2) a pure instantaneous burst, (3) a lower stellar mass-to-light
ratio (implicitly), (4) neglect of the difference between the mass of
stars formed over a Hubble time and that measured at $z=0$. In
addition, \citet{mg04} adopt a SNIa delay time distribution that is
both more sharply peaked and with a more prominent tail to long
delays.  Nevertheless, their results are broadly consistent with those
presented here in marginally allowing scenarios where $z(t_{\rm
form})=3$ and the mean delay time is 3 Gyr (Figure 12). The present
study goes further in explicitly considering a range of IMFs and
possible star formation and outflow histories to calculate the
evolution of ICM enrichment, and in concluding that models with more
prompt enrichment more accurately predict the flat ICM Fe evolution
that is observed.

\citet{po05} analyzed the cluster metal inventory at $z=0$. An
instantaneous burst of star formation was assumed, with
self-consistent treatment of the stellar mass-to-light ratio and mass
return, although the stellar Fe abundance was put in by hand. Since
chemical evolution is not modeled, the observed ICM Fe evolution is
not utilized to evaluate scenarios. Models with top-heavy IMFs or
non-standard SNIa parameters were not explicitly
considered. Nevertheless, their rejection of a standard IMF is robust
and in agreement with the conclusions presented here.

\citet{dkw} utilized N-body simulations in a $\Lambda$CDM universe,
conjoined with a parameterized semi-analytic treatment of
astrophysical processes in the baryonic component such as cooling,
star formation, feedback, and galactic winds, to simulate the chemical
evolution of the IGM and the ICM. Yields appropriate to a top-heavy
IMF are adopted, although mass return and the photometric properties
of galaxies were computed for a Salpeter IMF. In agreement with the
results of this work, they found that in models tuned to reproduce the
observational properties of $z=0$ galaxies, ICM enrichment was
invariably concentrated to high redshifts -- a direct result of the
$z=5$ peak in the star formation rate and the assumption of
instantaneous recycling.

The semi-analytic models of Nagashima et al. (2005a) included a more
detailed treatment of chemical evolution, and adopted some of the
enhancements listed in \S4.7 below, such as a two-phase ISM and
separate treatment of galaxy disk and bulge components. They fixed the
disk and bulge IMFs, with the latter characterized by a monotonic flat
slope (thus maximizing the SNII/SNIa ratio in this population), and
also adopt a single prescription for computing the SNIa rate. Neither
the star formation, nor the SNIa, rate histories were shown or
directly compared to observations; however, the latter seems to have a
mean delay intermediate between those most often considered here ({\it
i.e.}  $t_c=0.5$, 3 Gyr). Yet again, it was concluded that a top-heavy
IMF (corresponding to $\alpha_{3X}=1$) is required to produce
sufficient metals -- an inference further supported by comparing
Fe-to-$\alpha$-element ratios in individual elliptical galaxies with
those in galaxy formation simulations \citep{n05b}. ICM Fe enrichment
of the ICM occurs early in their models, as observed, with
$\alpha$-element enrichment ensuing even more promptly.

Contrary to the results of the present investigation, and those
described above, \cite{et05} concluded that ICM Fe can be accounted
for by enrichment from star formation with a Salpeter IMF and SNIa
with delay time and normalization as found in the field by \citet{s04}
and \citet{d04} -- similar to model 1N2.3 in the present work that was
dismissed as failing to produce adequate Fe. \cite{et05} acknowledged,
however, that not enough stars are observed (via $L_B$) for this
scenario to be fully coherent. This internal inconsistency can be
clarified as follows. The supernova parameters adopted in \citet{et05}
result in an Fe yield of $1.05\ 10^{-3}$ per solar mass of star
formation compared to the estimated total cluster Fe mass of $8.6\
10^{-3}$ per solar mass of ICM -- {\it i.e.} a mass ratio of
stars-formed to gas of 0.82, or a ratio of stars formed to stars
observed today of 6.3 (for a present-day cluster star-to-gas ratio of
0.12) is implied. The implied return fraction of 0.84 is grossly
inconsistent with the initial assumption of a Salpeter IMF.

\subsection{Model Shortcomings and Future Refinements}

One might be troubled by the high initial ISM baryon fractions, and
the high resulting galactic Fe abundances\footnote{Again, I note that
the models do not distinguish between metals locked up in stellar
remnants and those observationally accessible in still-living stars;
the latter can be significantly less metal-rich than the former.}, in
models that best explain the ICM Fe and SNIa observations. The need
for a reservoir to provide for the copious ensuing star formation (no
additional sources of star-forming gas -- via, e.g., infall -- are
subsequently introduced), and to soak up the newly synthesized metals
to be ejected into the IGM drives the requirement for the former. In
the context of the simple models constructed here, the latter is an
inevitable consequence of the early epoch for most star formation and
the observationally-driven requirement for early enrichment. A number
of refinements that may prove fruitful in follow-up studies, some of
which are relevant to these issues, may be considered.

In the models presented here, star formation and IGM-enriching
outflows originate in the same single-component ISM. One may subdivide
the ISM into hot and cold phases, with stars exclusively forming from
the latter and outflows exclusively originating in the former. The two
phases can exchange mass and be distinctly enriched by supernovae. One
may alternatively, or additionally, include ``bias'' parameters to
ensure that outflowing and/or star-forming gas is preferentially
metal-enriched relative to the average in the ISM.  These all may
affect the relationship between the required duration of galactic
outflows and the amount of metals locked up in stars. Likewise,
requirements on the magnitude and timing of outflows may be altered by
explicitly considering (re)accretion of IGM onto galaxies. Outflows
associated with SNII and SNIa may be ``decoupled'' by associating
distinct normalizations with each type.

The number of components may be further expanded in order to, e.g,
allow for separate treatment of bulges and disks within the stellar
component as regards star formation rate, IMF, SNIa normalization and
delay time distribution, outflows/inflows, etc. -- rather than the
less physically motivated division into ``field-like'' and
``cluster-enhancement'' components. This additional level of detail
clarifies the field/cluster dichotomy, and would enable a more
detailed evaluation of the star formation histories in the models.

A more sophisticated treatment of supernovae could include yields that
depend on progenitor mass, thus introducing an IMF-dependence for
average yields, or on the metallicity of the progenitor population.
These are often implemented for SNII, but could be considered for SNIa
as well, as could a metal dependence of the number of SNIa per star
formed.

A less empirical treatment of star formation could be implemented in a
global manner by specifying an explicit, parameterized dependence on
total ISM mass density constrained by observations of field star
formation rates and the distinct stellar inventories in clusters and
in the field.  An explicit metallicity-dependent IMF could be
considered. The formation and evolution of binary star systems may be
distinctly and explicitly traced, and granted a role in determining
the SNIa rate.

Population III enrichment may be more carefully considered, especially
in relation to metal build-up at high-redshift. The overall impact of
such a component is generally considered minor in the IGM out to
relatively high redshift (Yoshida, Bromm \& Hernquist 2004; Norman,
O'Shea, \& Paschos 2004; but, see Salvaterra \& Ferrara 2003), but may
be of enhanced significance in clusters \citep{l01}.

Finally, one should consider the abundance evolution of additional
elements, such as O, Mg, and Si to further constrain the combination
of IMF, relative SNII/SNIa normalization, and outflow parameters, as
well as the average yields themselves.  

While the above considerations indicate limitations of the present
work, the physically and observationally motivated, self-consistent
models constructed here reveal a number of implications and
puzzles. By virtue of the simplicity and limited number of free model
parameters in this approach, one can draw conclusions and identify
persistent paradoxes -- summarized in the section that follows -- that
are qualitatively model-independent and easily traced to particular
phenomenon or assumptions.

\section{Summary and Concluding Remarks}

I constructed models for the coupled mass density and Fe abundance
evolution of stars, interstellar (potentially star-forming) gas, and
intergalactic (inert) gas in a closed box, with application to the
universe as a whole and extension to the special environments of rich
clusters of galaxies. I focused on Fe because of its power as a
diagnostic of SNIa astrophysics and the unmatched constraints provided
by recent observations of the evolution of the SNIa rate and ICM Fe
abundance. I adopted empirically based source terms to mitigate the
limitations in our understanding of star formation and the physics and
astrophysics of supernova explosions. I considered the effects of
varying the star formation rate history, initial mass function, SNII
Fe yields, SNIa rate normalization and distribution of delay times,
and strength and duration of galactic outflow on field and cluster
observables. Confrontation of models with observations of the three
components in both environments, and over a range in redshift,
constrain the mechanisms, and highlight the distinctions, of metal
enrichment in the field and in clusters that depend on the physical
characteristics of star formation and SNIa explosions.

\subsection{Chemical Evolution of Fe in the Universe}

The average star formation history in the universe is measured out to
redshift $z\sim 6$ \citep{g04}, and a SNIa delay time distribution
function derived from a comparison with the SNIa rate history to
$z>1$. The results of direct application of these functions --
assuming a standard IMF, and standard SNIa and SNII Fe yields -- to
chemical evolutionary models are consistent with metallicities
measured in damped Ly$\alpha$ systems out to $z\sim 3.5$ and in the
IGM to $z\sim 5$. A standard IMF also correctly predicts measurements
of the buildup of stellar mass over cosmic time, and estimated SNII
rates to $z\sim 1$. A careful accounting of star formation, stellar
mass return, and galactic winds implies that the ISM density is $>3$
times that in damped Ly$\alpha$ systems (DLAs) at $z=2.5$, supporting
suggestions that the ``missing metal'' problem is actually a ``missing
gas'' problem -- {\it i.e.} that there is more potentially
star-forming gas than currently measured in DLAs. The delay in SNIa
enrichment implies that less than 10\% of the Fe in field galaxies was
in place at $z=2$ and that the Fe yield per star formed at that
redshift was approximately half the current value. One expects a
negative correlation of the Fe-to-$\alpha$ elemental abundance ratio
with stellar population age. Although some field star formation may
proceed with a flat IMF \citep{v06}, drastic departures from the
standard model presented here, unless constructed in a highly
contrived manner, are ruled out by observations and their mutual
consistency within the standard framework.

\subsection{Chemical Evolution of Fe in Clusters}

The amount of Fe per baryon measured in clusters exceeds that
predicted by the successful and internally consistent field model
described above by more than a factor of four. And yet, surveys fail
to detect supernovae -- in particular SNIa that most efficiently
synthesize Fe -- in sufficient numbers to account for this
hyper-enrichment.  In this paper I presented, and evaluated the
plausibility and implications of, possible resolutions to this now
well-established (Portinari et al. 2005, and references therein)
paradox.

Clusters are typically inferred to have roughly twice the stellar mass
fraction as in the field, and cluster galaxies are believed to form
stars more rapidly. However, these factors are insufficient (although
the latter is necessary) for reproducing the observed Fe abundance and
its mild evolution since $z\sim 1$. Fe must be synthesized in
abundance at an early epoch, and must be efficiently transported from
galaxies to the ICM. Therefore, a significant fraction of star
formation must proceed rapidly with a top-heavy IMF (see, also,
Finoguenov, Burkert, \& B\"ohringer 2003). Star formation is $3-5$
times more efficient in rich clusters than in the field, mitigating
the overcooling problem in numerical cluster simulations. Both the
fraction of baryons cycled through stars, and the fraction of the
total present-day stellar mass in the form of stellar remnants, are
substantially greater in clusters than in the field.  Assuming that
metals are well-mixed in the ISM before ejection, $\sim 40-70$\% of
the ICM must originate in galaxies (see, also, Moretti et al. 2003),
with the lower values generally associated with galactic outflows
assumed to continue for longer durations.

Confirming and extending the conclusions drawn by previous studies
(see \S 4.6), I demonstrated that the observed mild amount of
evolution in Fe is most accurately reproduced in models where the
enrichment tracks star formation, implying that synthesis of cluster
Fe was dominated by SNII (if the SNIa rate normalization is scaled
down from its value in the field according to the relative number of
$3-8~{\rm M}_{\odot}$ stars) and/or SNIa with short delay times whose
progenitors originated during a phase of rapid, top-heavy star
formation (see, also, Scannapieco \& Bildsten 2005). In the former,
$\ge 0.1~{\rm M}_{\odot}$ per SNII is required; in the latter $\ge
10^{-3}~{\rm M}_{\odot}$ of Fe from SNIa per star formed is required.

The best matches to the low observed $z<1$ cluster SNIa rates are
reproduced in models where cluster star formation does not exclusively
occur at high redshift, but continues to at least $z\sim 0.5$. This is
a more prolonged history of star formation than inferred in the
central regions of rich clusters, and is contingent on the assumption
of a steep decline in the SNIa delay time distribution function beyond
3 Gyr.

Observations of high Fe content in the ICM and its persistence to
$z=1$, in combination with constraints on SNIa enrichment from rates
measured to similar redshifts, confirms the rapid and efficient nature
of star formation in galactic spheroids in clusters, and provides
overwhelming evidence that this star formation was characterized by a
top-heavy IMF that produced Type Ia less efficiently and/or with
shorter delay times (on average) than in field galaxies. Rich galaxy
clusters are, indeed, special environments and should only be treated
as ``fair samples of the universe'' in a limited and well-defined
context.

\acknowledgments
I am grateful to an anonymous referee for carefully considered and
constructive comments on the original draft of this manuscript.


\clearpage
\setlength{\textheight}{9in}
\begin{deluxetable}{cccccccccccc}
\tablewidth{0pt} \tablecaption{Model Input Parameters} 
\tablehead{
\colhead{model} & \colhead{$f_{\rm stars}$\tablenotemark{a}} &
\colhead{$\alpha_3$} & \colhead{$f_{\rm Xstars}$\tablenotemark{b}} & 
\colhead{$m_{DX}$} & \colhead{$\alpha_{3X}$} & 
\colhead{${{t_{\rm wind}}\over {t_{\rm now}}}$} & 
\colhead{${{\Delta t_{\rm wind}}\over {t_{\rm now}-t_{\rm form}}}$} & 
\colhead{$f_{\rm ISM}({t_{\rm form}})$\tablenotemark{c}} &
\colhead{$t_c$\tablenotemark{d}} &
\colhead{$t_{cX}$\tablenotemark{e}} 
}
\startdata 
1N2.3 & 0.059 & 2.3 & 0 & \nodata & \nodata & \nodata & \nodata & 0.35
& 3.0 & \nodata\\
1N1.05Wc & 0.059 & 1.05 & 0 & \nodata & \nodata & \nodata & 0.5 & 0.38
& 3.0 & \nodata\\
1H0.8Wc & 0.029 & 1.8 & 0.029 & 0.5 & 0.8 & \nodata & 0.5 & 0.38 & 3.0
& 3.0\\
1R1.05Wc & 0 & \nodata & 0.059 & 2.0 & 1.05 & \nodata & 1.0 & 0.32 &
3.0 & 3.0\\
2H0.8Wc & 0.059 & 2.3 & 0.059 & 8.0 & 0.8 & \nodata & 0.66 & 0.34 &
3.0 & 3.0\\
2H1.05 & 0.059 & 2.3 & 0.059 & 2.0 & 1.05 & \nodata & \nodata & 0.48 &
3.0 & 3.0\\
2H1.05Wx & 0.059 & 2.3 & 0.059 & 2.0 & 1.05 & 1.0 & \nodata & 0.52 &
3.0 & 3.0\\
2H1.3Wx & 0.059 & 2.3 & 0.059 & 2.0 & 1.3 & 0.5 & \nodata & 0.51 & 3.0
& 3.0\\
2H1.3Wc & 0.059 & 2.3 & 0.059 & 2.0 & 1.3 & \nodata & 1.0 & 0.34 & 3.0
& 3.0\\
2H1.55Wx & 0.059 & 2.3 & 0.059 & 0.5 & 1.55 & 1.0 & \nodata & 0.68 &
3.0 & 3.0\\
2R1.55Wx & 0 & \nodata & 0.12 & 2.0 & 1.55 & 1.0 & \nodata & 0.50 &
\nodata & 3.0\\
\hline
2H1.05WxSt1 & 0.059 & 2.3 & 0.059 & 2.0 & 1.05 & 1.0 & \nodata & 0.51
& 0.5 & 0.5\\
2H1.05WxSt2 & 0.059 & 2.3 & 0.059 & 2.0 & 1.05 & 1.0 & \nodata & 0.51
& 3.0 & 0.5\\
2R1.55WxSt & 0 & \nodata & 0.12 & 2.0 & 1.55 & 1.0 & \nodata & 0.50 &
\nodata & 0.5\\
2R1.55WxSz\tablenotemark{f} & 0 & \nodata & 0.12 & 2.0 & 1.55 & 1.0 &
\nodata & 0.83 & \nodata & 3.0\\
2R1.55WxStpz\tablenotemark{f} & 0 & \nodata & 0.12 & 2.0 & 1.55 & 1.0
& \nodata & 0.75 & \nodata & 1.5\\
2R1.55WxStz\tablenotemark{f} & 0 & \nodata & 0.12 & 2.0 & 1.55 & 1.0 &
\nodata & 0.58 & \nodata & 0.5\\ 
2H1.05WxSn\tablenotemark{gh} & 0.059 & 2.3 & 0.059 & 2.0 & 1.05 & 1.0
& \nodata & 0.60 & 3.0 & 3.0\\
2H1.05WxStn\tablenotemark{gh} & 0.059 & 2.3 & 0.059 & 2.0 & 1.05 & 1.0
& \nodata & 0.60 & 3.0 & 0.5\\
2R1.55Stnz\tablenotemark{fgh} & 0 & \nodata & 0.12 & 2.0 & 1.55 & \nodata
& \nodata & 0.73 & \nodata & 0.5\\
\enddata
\tablenotetext{a}{Present-day baryon fraction in stars formed in
normal mode.}
\tablenotetext{b}{Present-day baryon fraction in stars formed in rapid
mode.}
\tablenotetext{c}{Initial baryon fraction in star-forming gas (ISM).}
\tablenotetext{d}{Mean SNIa delay, in Gyr, for normal star formation mode.}
\tablenotetext{e}{Mean SNIa delay, in Gyr, for rapid star formation mode.}
\tablenotetext{f}{Formation epoch of the first Population II stars
moved forward from $z=10$ to $z=3$.}
\tablenotetext{g}{SNIa normalization reduced according to IMF.}
\tablenotetext{h}{SNII yield increased from 0.07 to $0.1~{\rm
M}_{\odot}$.}
\end{deluxetable}
\clearpage
\setlength{\textheight}{8.4in}
\begin{deluxetable}{cccccccc}
\tablewidth{0pt}
\tablecaption{Model Characteristics and Results at $z=0$}
\tablehead{
\colhead{model} & \colhead{$\dot {n_{\rm SNII}}$} & 
\colhead{$\dot {n_{\rm SNIa}}$} & \colhead{$f_{\rm Ia}{\rm(Fe)}$} & 
\colhead{${y^{\rm Fe}}_{\rm stars}$} &
\colhead{${\langle{\rm Fe}\rangle}_{\rm bar}$} & 
\colhead{${\langle{\rm Fe}\rangle}_{\rm IGM}$} & 
\colhead{${\langle{\rm Fe}\rangle}_{\rm gal}$}
}
\startdata
1N2.3 & 0.59 & 0.25 & 0.58 & 1.3 & 0.13 & 0.056 & 1.0\\
1N1.05Wc & 3.7 & 0.84 & 0.42 & 1.8 & 0.59  & 0.41 & 2.8\\
1H0.8Wc & 0.76 & 0.19 & 0.45 & 1.8 & 0.51 & 0.39 & 1.9\\
1R1.05Wc & 0.019 & 0.015 & 0.44 & 1.8 & 0.61 & 0.43 & 2.8\\
2H0.8Wc & 0.60 & 0.26 & 0.53 & 1.5 & 0.50 & 0.38 & 1.3\\
2H1.05 & 0.61 & 0.26 & 0.46 & 1.7 & 0.74 & 0.26 & 3.8\\
2H1.05Wx & 0.61 & 0.26 & 0.46 & 1.7 & 0.74 & 0.36 & 3.2\\
2H1.3Wx & 0.61 & 0.26 & 0.45 & 1.8 & 0.62 & 0.40 & 2.1\\
2H1.3Wc & 0.61 & 0.26 & 0.45 & 1.8 & 0.62 & 0.42 & 1.9\\
2H1.55Wx & 0.61 & 0.26 & 0.43 & 1.7 & 0.66 & 0.35 & 2.6\\
2R1.55Wx & 0.023 & 0.017 & 0.43 & 1.9 & 0.73 & 0.37 & 3.0\\
\hline
2H1.05WxSt1 & 0.61 & 0.099 & 0.46 & 1.7 & 0.74 & 0.40 & 2.9\\
2H1.05WxSt2 & 0.61 & 0.25 & 0.46 & 1.7 & 0.74 & 0.40 & 2.9\\
2R1.55WxSt & 0.023 & 0.0024 & 0.43 & 1.9 & 0.73 & 0.39 & 2.9\\
2R1.55WxSz & 0.077 & 0.058 & 0.43 & 1.9 & 0.72 & 0.40 & 2.8\\
2R1.55WxStpz & 0.077 & 0.018 & 0.43 & 1.9 & 0.73 & 0.40 & 2.8\\
2R1.55WxStz & 0.077 & 0.0083 & 0.43 & 1.9 & 0.73 & 0.40 & 2.8\\
2H1.05WxSn & 0.61 & 0.025 & 0.20 & 1.7 & 0.71 & 0.40 & 2.7\\
2H1.05WxStn & 0.61 & 0.025 & 0.20 & 1.7 & 0.71 & 0.41 & 2.7\\
2R1.55Stnz & 0.077 & 0.0055 & 0.25 & 2.1 & 0.80 & 0.40 & 3.4\\
\enddata

\tablecomments{Supernova rates, $\dot {n_{\rm SNII}}$ and $\dot
{n_{\rm SNIa}}$ are in units of $10^{-4}$ $Mpc^{-3}
yr^{-1}\delta^{-1}$, where $\delta$ is the baryon overdensity; $f_{\rm
Ia}{\rm(Fe)}$ is the fraction of Fe originating from SNIa, ${y^{\rm
Fe}}_{\rm stars}$ the Fe yield per star relative to the solar mass
fraction; ${\langle{\rm Fe}\rangle}_{\rm bar}$ the Fe produced per
total baryon; ${\langle{\rm Fe}\rangle}_{\rm IGM}$ and ${\langle{\rm
Fe}\rangle}_{\rm gal}$ the average Fe abundances in the ICM/IGM and in
galaxies, respectively.}

\end{deluxetable}

\clearpage

\begin{figure}
\plotone{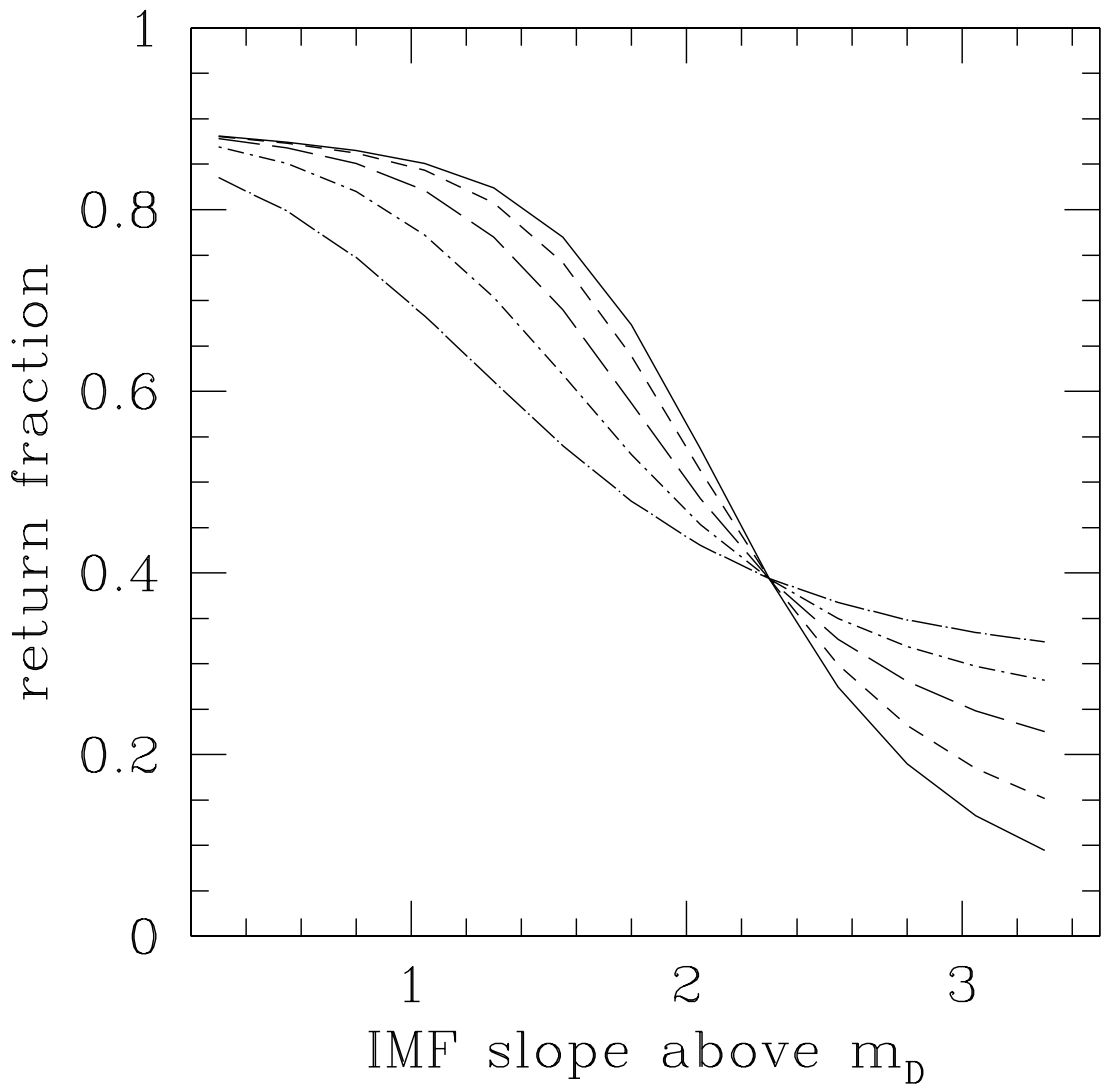}
\caption{Integrated return fraction as a function of IMF slope above
$m_D$ for $m_D=0.5$, 1, 2, 4, and $8~{\rm M}_{\odot}$ (solid,
short-dash, long-dash, dot--short-dash, and dot--long-dash line-type,
respectively). \label{fig1}}
\end{figure}

\clearpage

\begin{figure}
\plotone{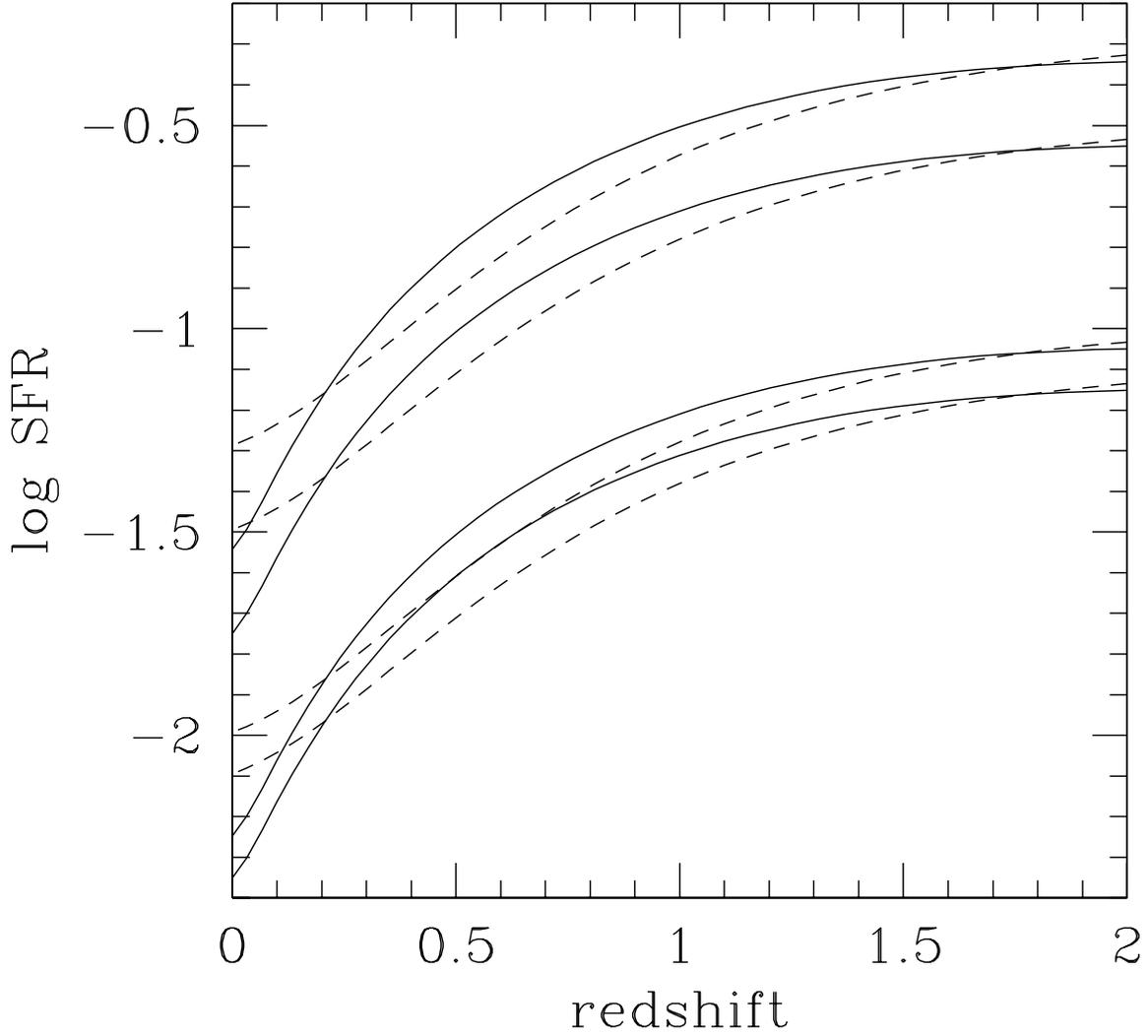}
\caption{\citet{b05} (solid curves) and \citet{s04} (broken curves)
star-formation rate (in ${\rm M}_{\odot}~{\rm Mp}c^{-3}~{\rm
yr}^{-1}$) history parameterizations for $m_D=1~{\rm M}_{\odot}$ and
(top to bottom) $\alpha_2=0.3$, 1.3, 2.3, and 2.8. The rates are
renormalized to yield the measured present-day stellar density when
integrated over time. \citep{fp04}. \label{fig2}}
\end{figure}

\clearpage

\begin{figure}
\plottwo{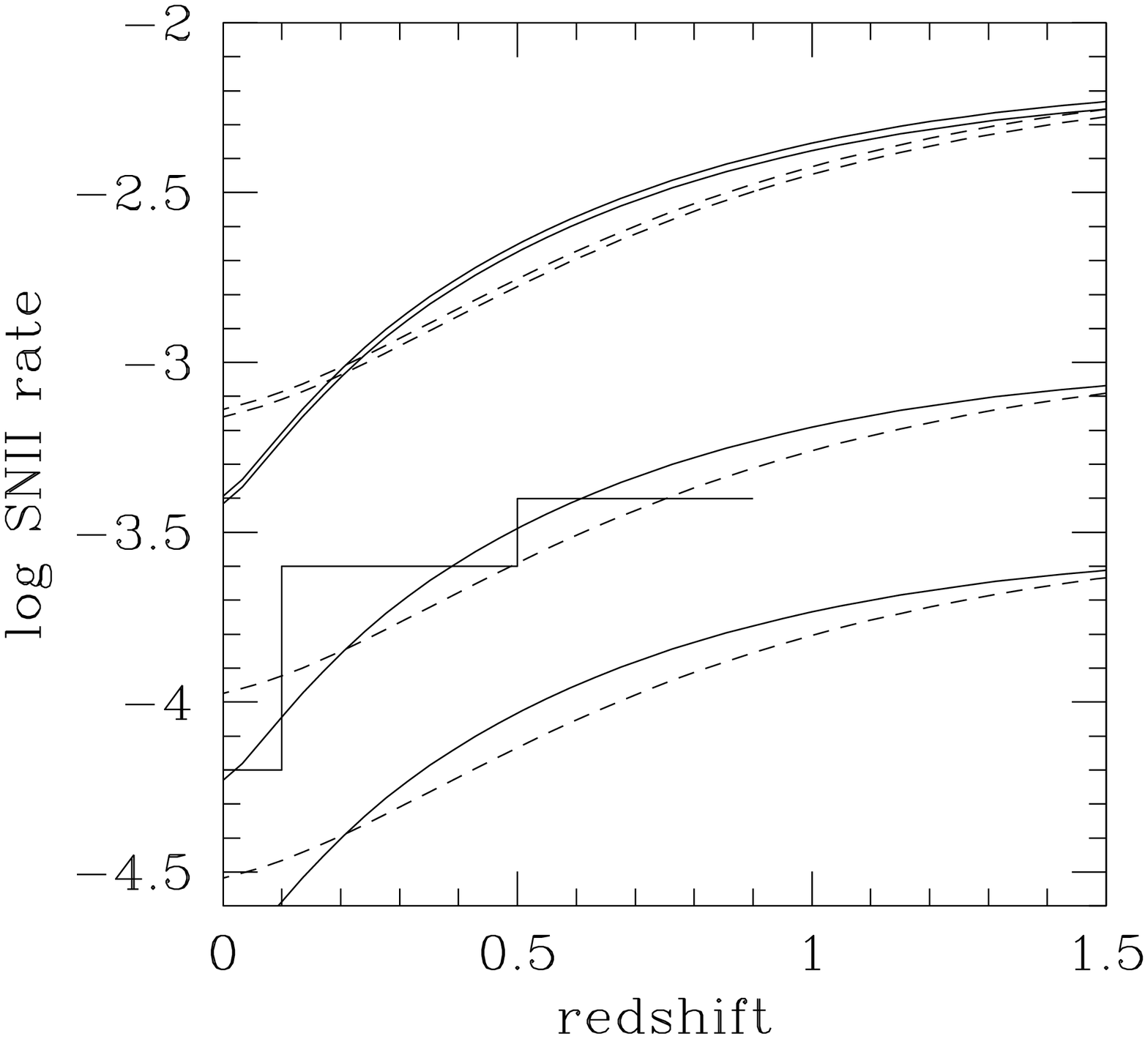}{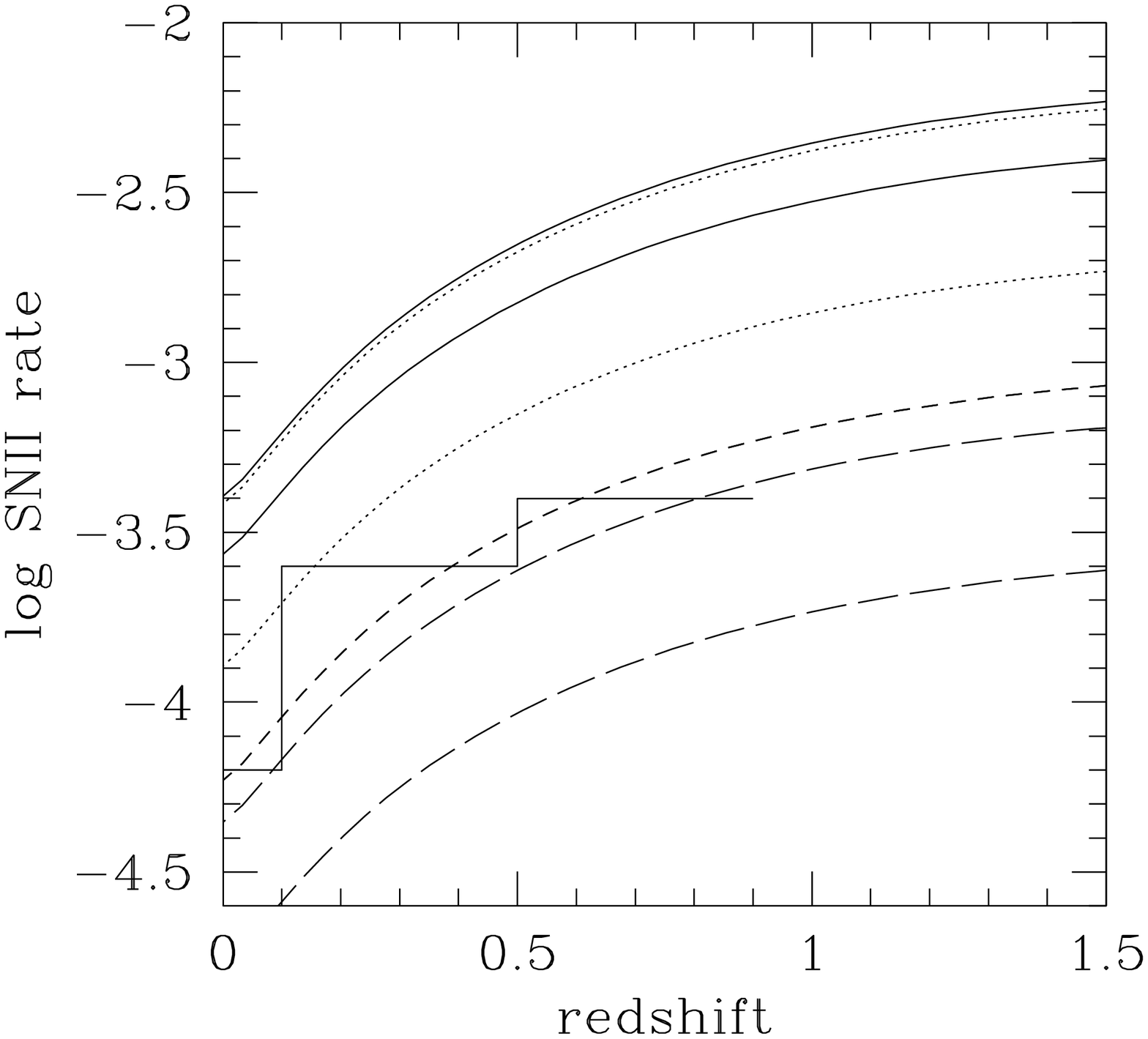}
\caption{left(a): Type II supernova rate (in ${\rm Mp}c^{-3}~{\rm
yr}^{-1}$) evolution corresponding to the curves in Figure
2. right(b): Type II supernova rate evolution for \citet{b05}
star-formation history parameterizations for $\alpha_2=0.3, 1.3, 2.3$,
and 2.8 (solid, dot, short-dash, and long-dash line-type
respectively). For $\alpha_2=0.3$ and 1.3, the upper (lower) curves
denote $m_D=1~{\rm M}_{\odot}$ ($m_D=8~{\rm M}_{\odot}$); this is
reversed for $\alpha_2=2.8$. The histogram shows the observed SNII
rates from \citet{d04} for the redshift intervals $z=0.1-0.5$ and
$0.5-0.9$ and the field rate from \citet{cet} for $z<0.1$ (converted
using a mass-to-blue-light ratio of 2.4; Fukugita \& Peebles 2004);
estimated uncertainties are at the $\sim 50$\% level.\label{fig3}}
\end{figure}

\begin{figure}
\plottwo{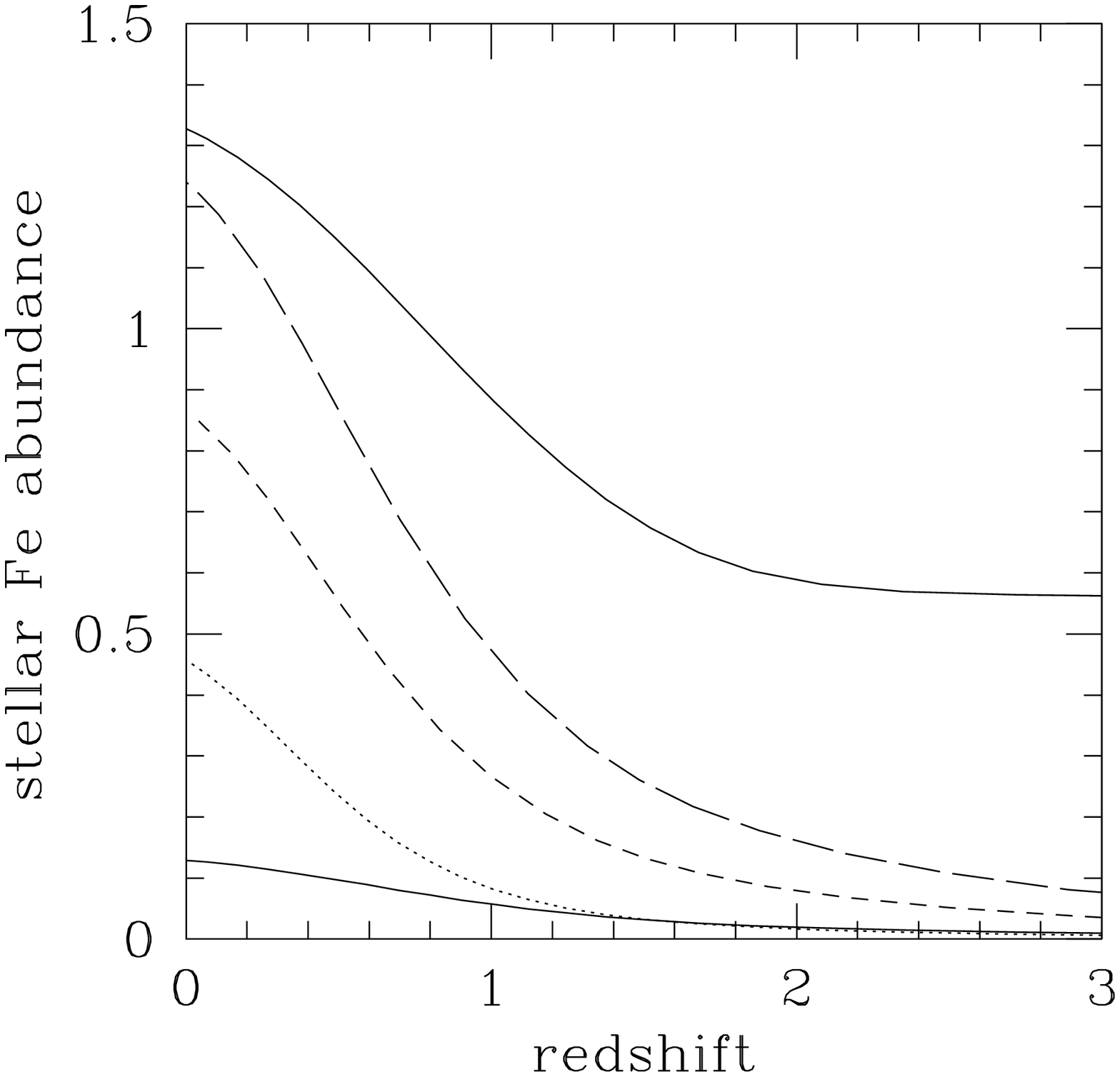}{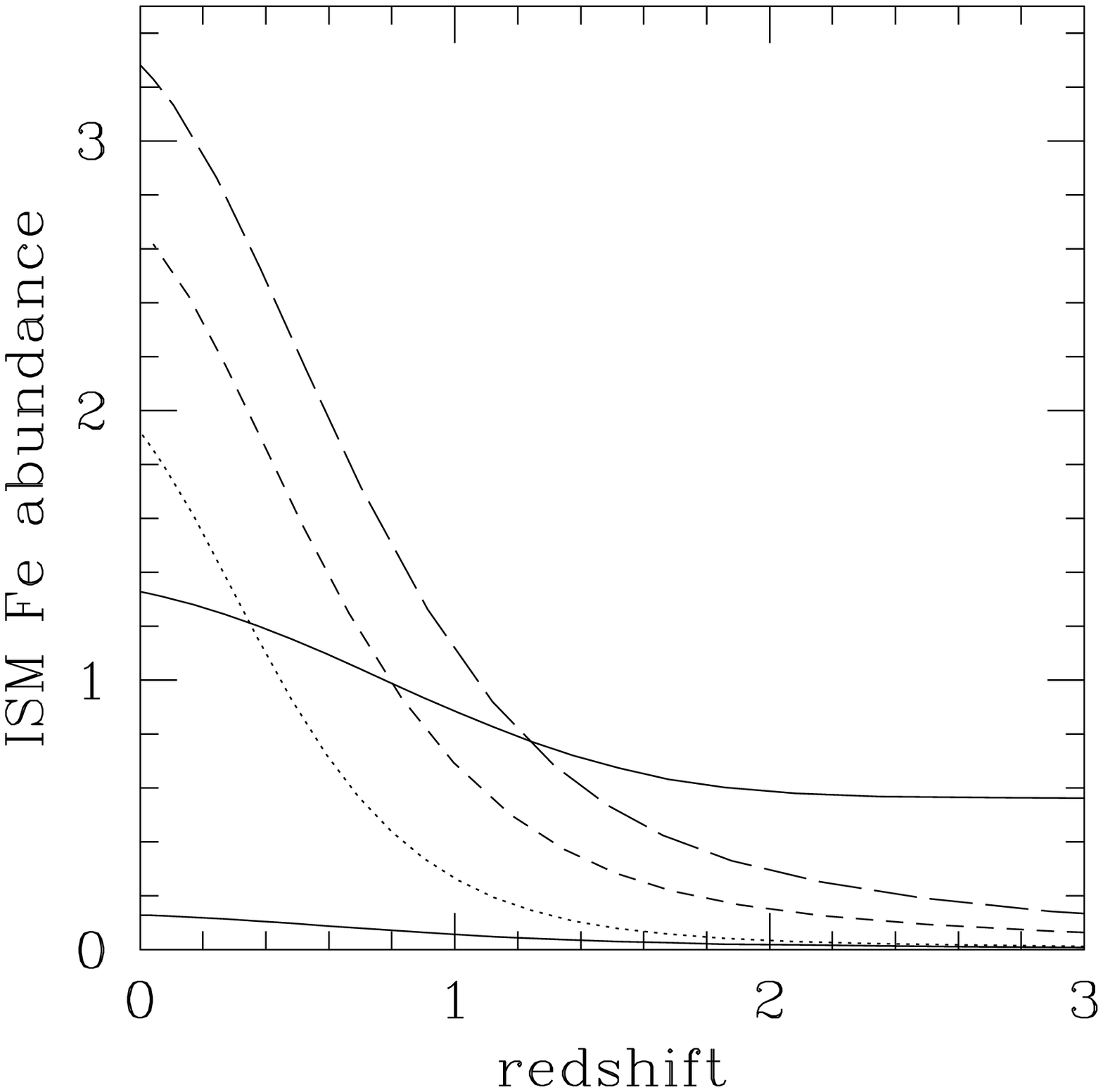}
\caption{left(a): Evolution of stellar Fe abundance for the standard
(see text) model. The cases with no wind ($\Omega_{\rm ISM}({t_{\rm
form}})\equiv \rho_{\rm ISM}({t_{\rm form}})/\rho_{\rm
crit}=\Omega_{\rm stars}({t_{\rm now}})+\Omega_{\rm ISM}({t_{\rm
now}})=0.0034$), maximum wind ($\Omega_{\rm ISM}({t_{\rm
form}})=\Omega_b-\Omega_{\rm III}\approx 0.045$) and an intermediate
case ($\Omega_{\rm ISM}({t_{\rm form}})=0.0076$) are denoted by
long-dash, dot, and short-dash line-type, respectively.  The upper
solid curve shows the Fe yield per unit star formed (relative to the
solar Fe mass fraction), the lower solid curve the overall average
abundance (or, equivalently, yield per baryon). right(b): Same as (a)
for the ISM Fe abundance. \label{fig4}}
\end{figure}

\clearpage

\begin{figure}
\plotone{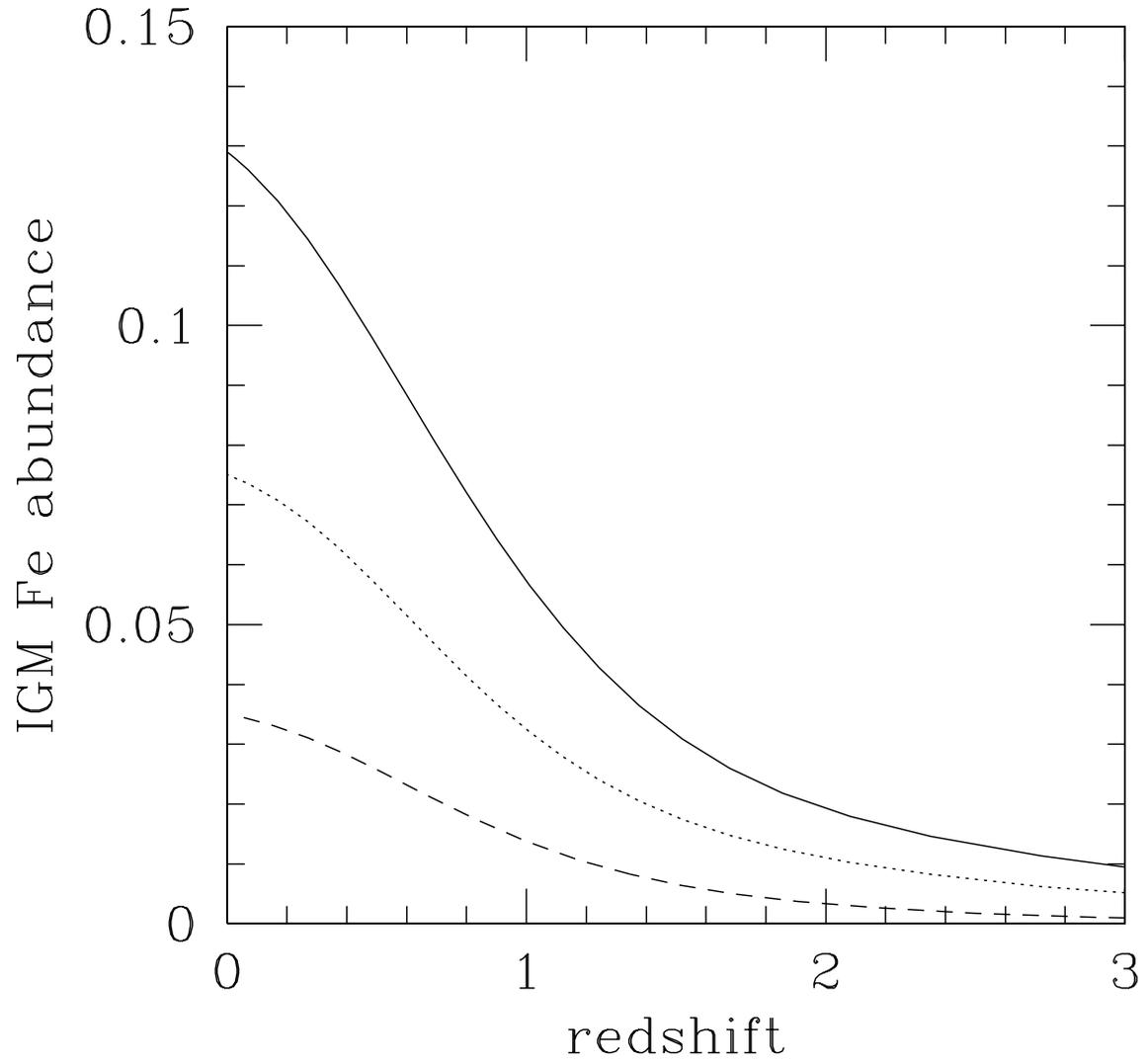}
\caption{Same as Figure 4 for the IGM Fe abundance (the abundance is 0
for the no-wind case).\label{fig5}}
\end{figure}

\begin{figure}
\plotone{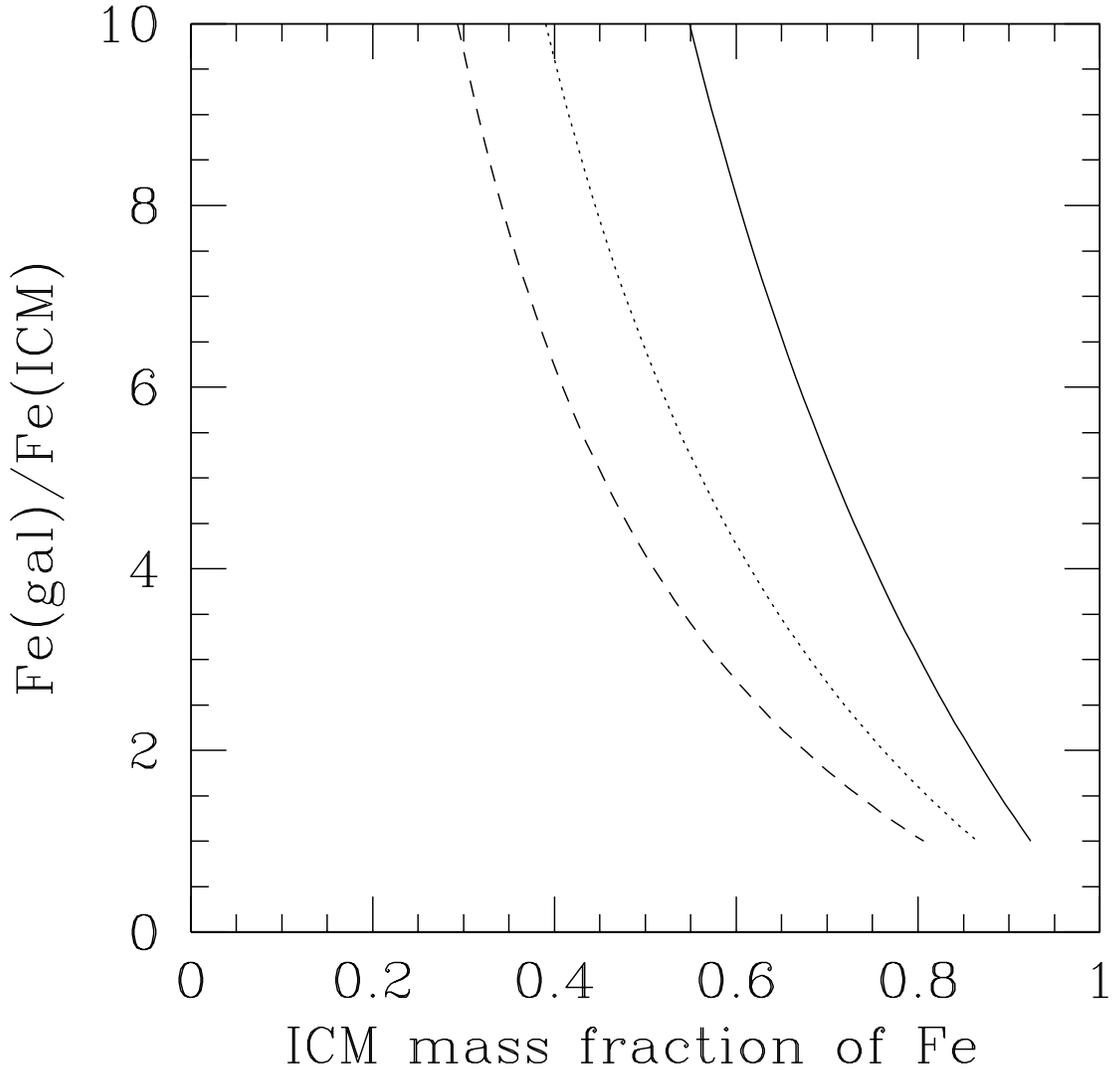}
\caption{Ratio of galactic Fe abundance to ICM Fe abundance as a
function of the present-day mass fraction of Fe in the ICM for overall
ICM mass fractions 0.924 (solid curve), 0.865 (dotted curve), and
0.806 (dashed curve) -- corresponding to baryon fractions in stars
equal to the universal value, twice the universal value, and three
times the universal value, respectively (the universal value for the
ISM baryon fraction is assumed).\label{fig6}}
\end{figure}

\clearpage

\begin{figure}
\plottwo{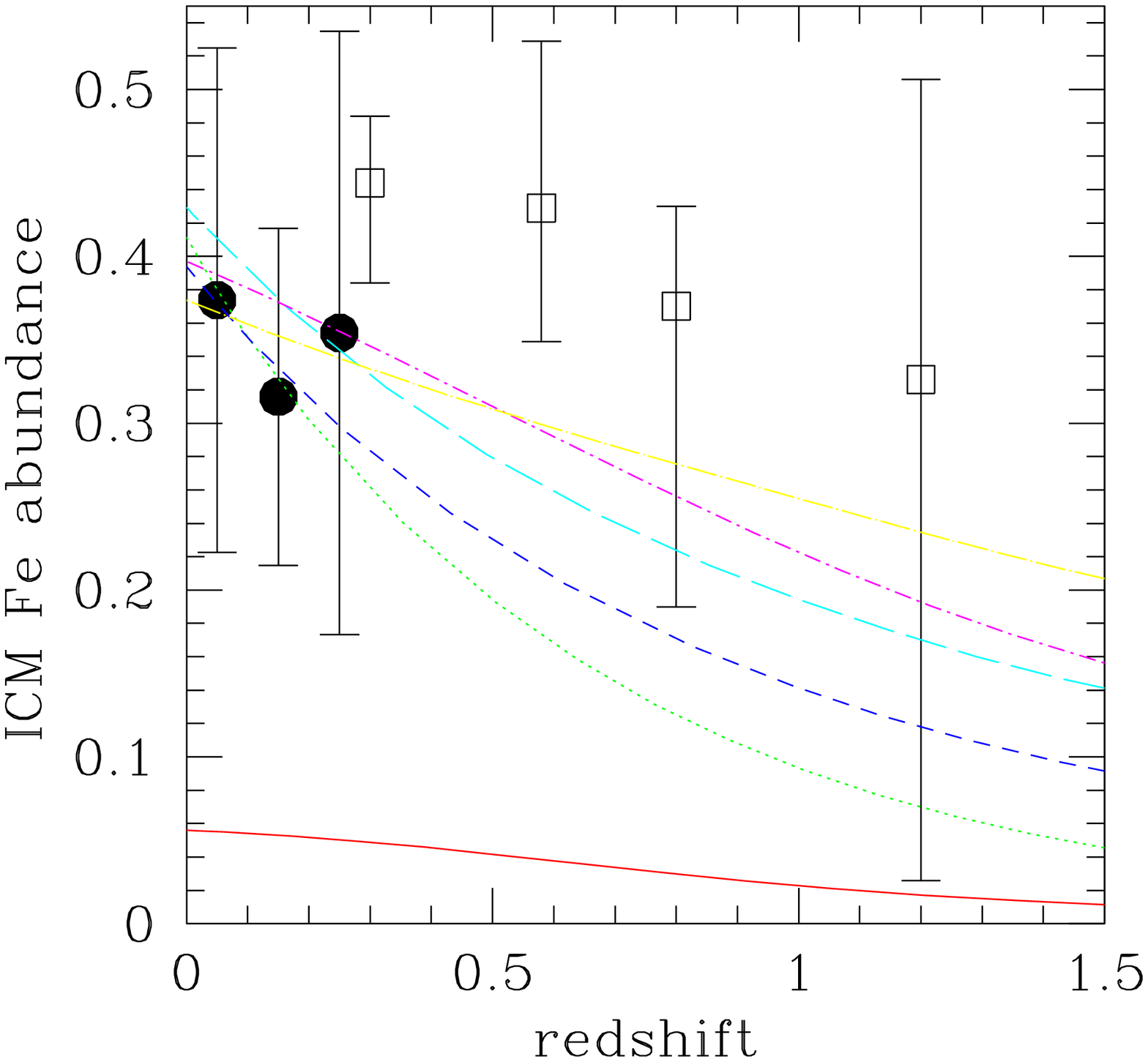}{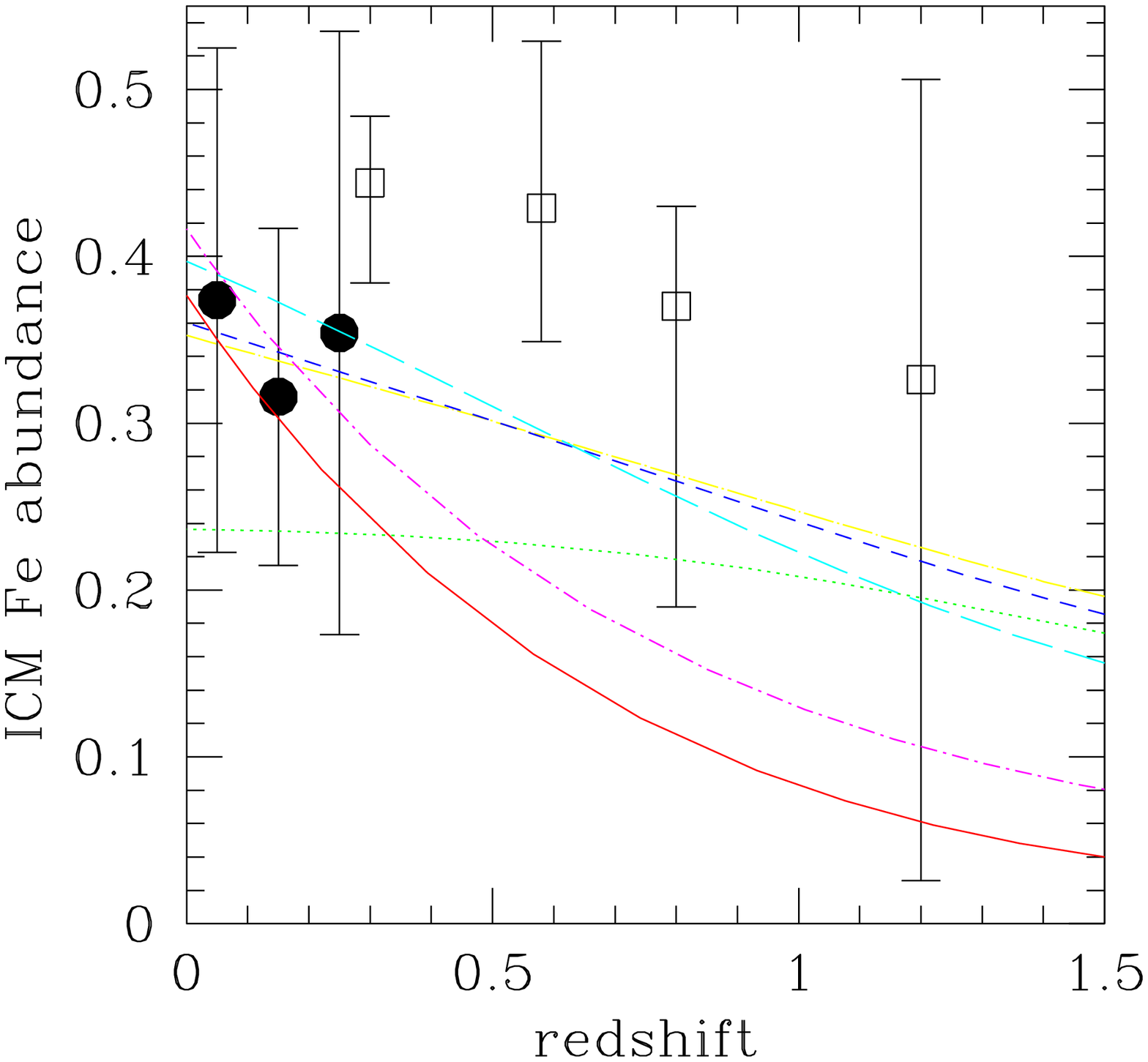}
\caption{left(a): Evolution of ICM Fe abundance for Models 1N2.3,
1N1.05Wc, 1H0.8Wc, 1R1.05Wc, 2H1.3Wx, and 2R1.55Wx (solid, dot,
short-dash, long-dash, dot-short-dash, dot-long-dash line-type
respectively). right(b): Evolution of ICM Fe abundance for Models
2H0.8Wc, 2H1.05, 2H1.05Wx, 2H1.3Wx, 2H1.3Wc, and 2H1.55Wx (solid, dot,
short-dash, long-dash, dot-short-dash, dot-long-dash line-type,
respectively). Filled circles denote abundances derived from {\it
ASCA} data of $kT>2$ keV clusters from \citet{bau05} with 90\%
confidence uncertainties, open squares those from {\it XMM-Newton}
data of $kT>5$ keV clusters from \citet{t03} with $2\sigma$
uncertainties.  See \S 3 and Tables 1 and 2 for model
details.\label{fig7}}
\end{figure}

\clearpage

\begin{figure}
\plottwo{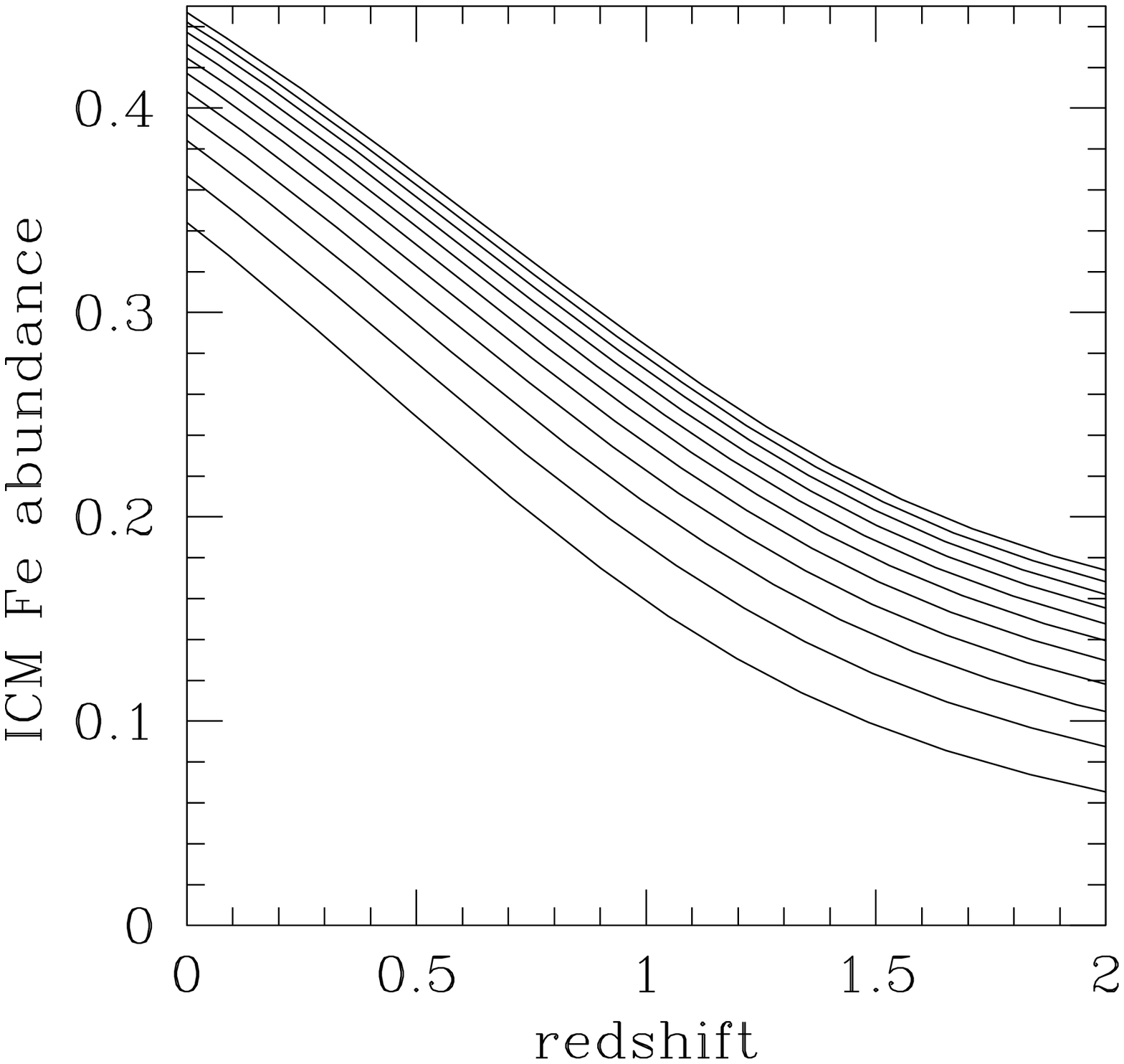}{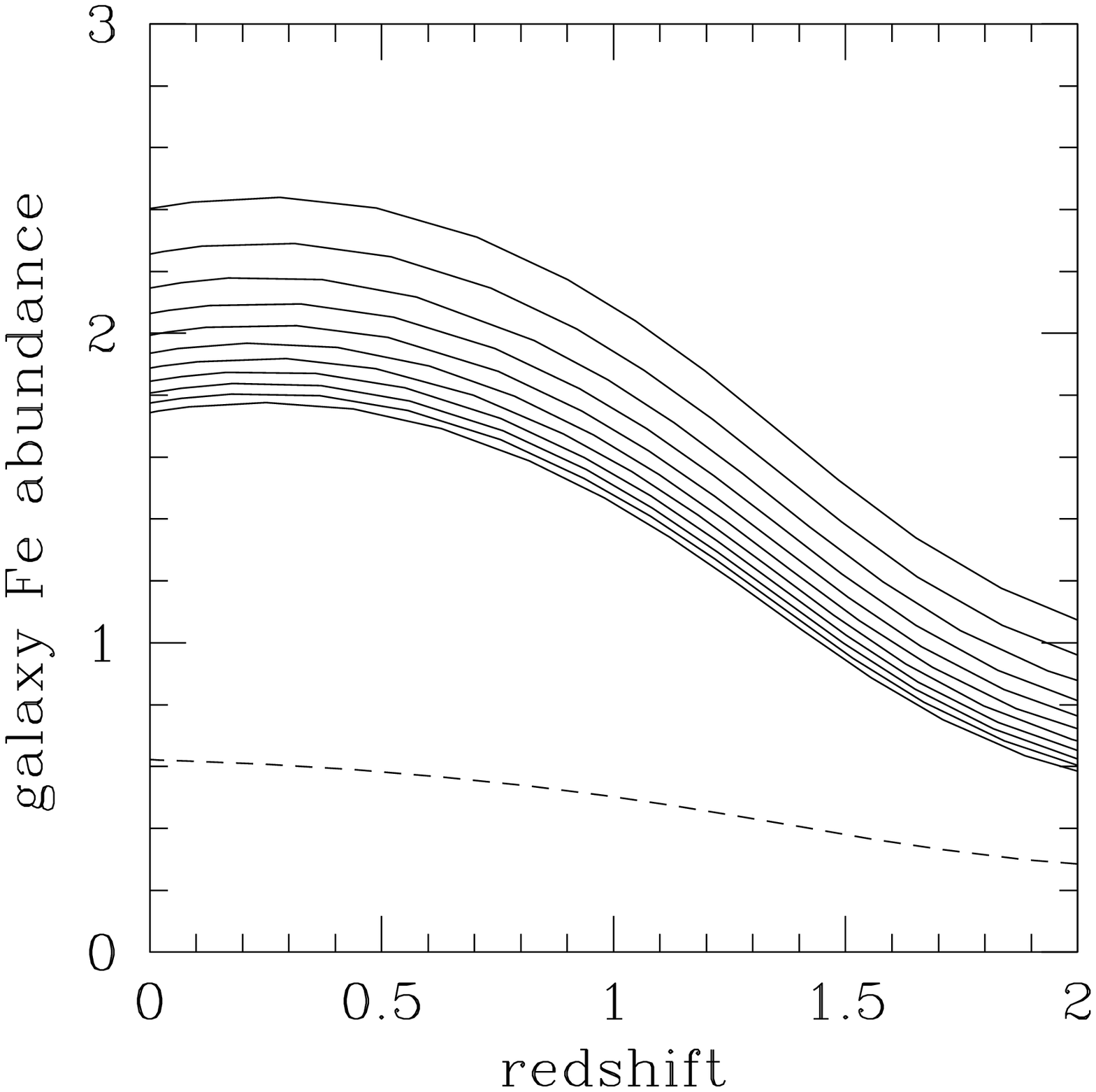}
\caption{left(a): Evolution of ICM Fe abundance for Model 2H1.3Wx with
variations in $K_{\rm GW}$ ranging from its minimum to maximum allowed
value. These correspond to initial ISM baryon fractions, $f_{\rm
ISM}({t_{\rm form}})=0.30$, 0.37, 0.44, 0.51, 0.58, 0.65, 0.72, 0.79,
0.86, 0.93, and 1.0 (lower to upper curves). The $f_{\rm ISM}({t_{\rm
form}})=0.51$ curve is also plotted in Figures 7a and 7b. right(b):
Same as (a) for galactic abundance (with lower curves corresponding to
smaller values of $f_{\rm ISM}({t_{\rm form}})$). The stellar Fe
abundance evolution is shown by the broken line.\label{fig8}}
\end{figure}

\clearpage

\begin{figure}
\plotone{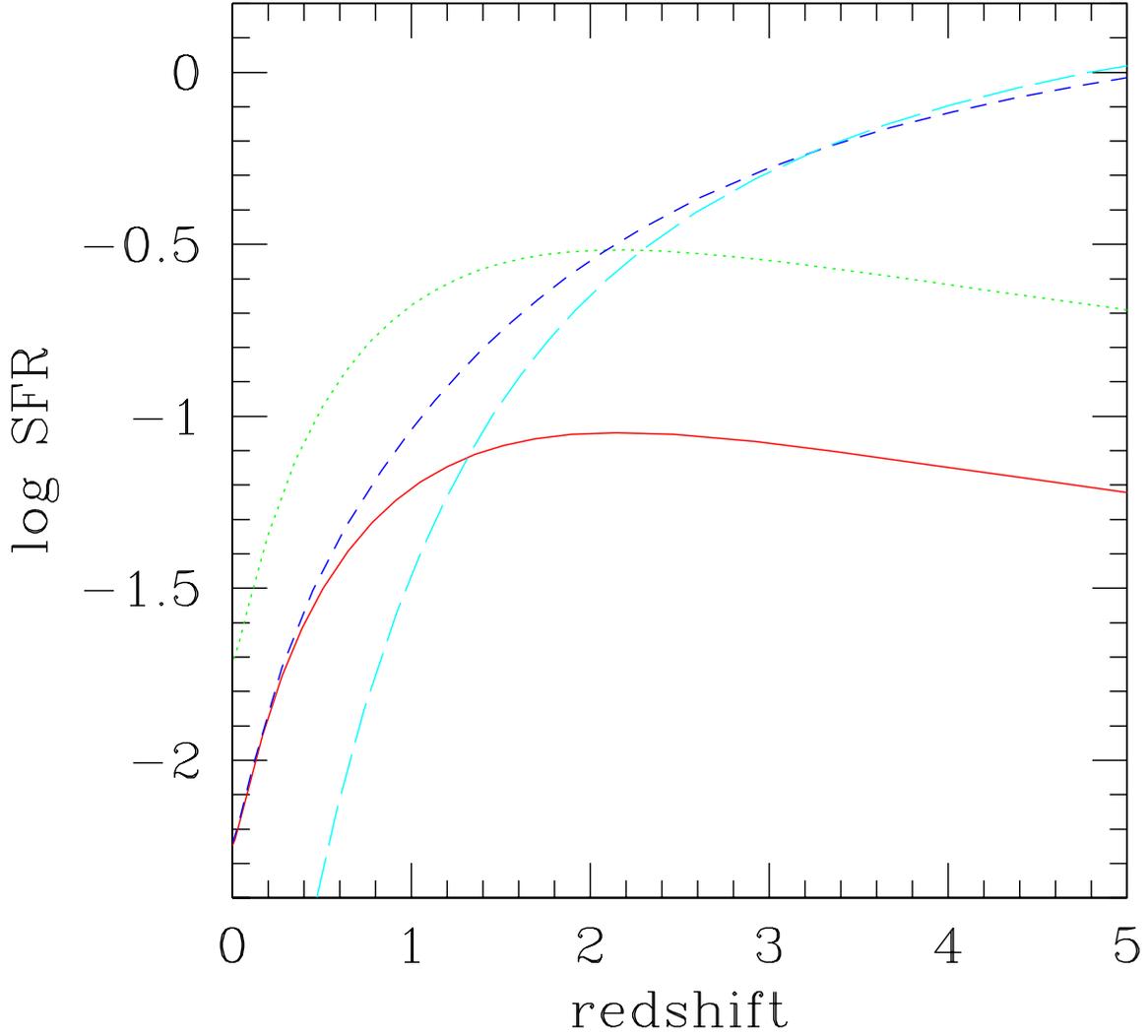} 
\caption{Star formation histories for Models 1N2.3, 1N1.05Wc, 2H1.3Wx,
and 2R1.55Wx (solid, dot, short-dash, long-dash line-types,
respectively.) Star formation rates are expressed in units of
$\delta^{-1}~{\rm M}_{\odot}~{\rm Mp}c^{-3}~{\rm yr}^{-1}$ where
$\delta$ is the baryon overdensity. The Model 1N2.3 curve matches the
(scaled) observed SFH for field galaxies.\label{fig9}}
\end{figure}

\clearpage

\begin{figure}
\plottwo{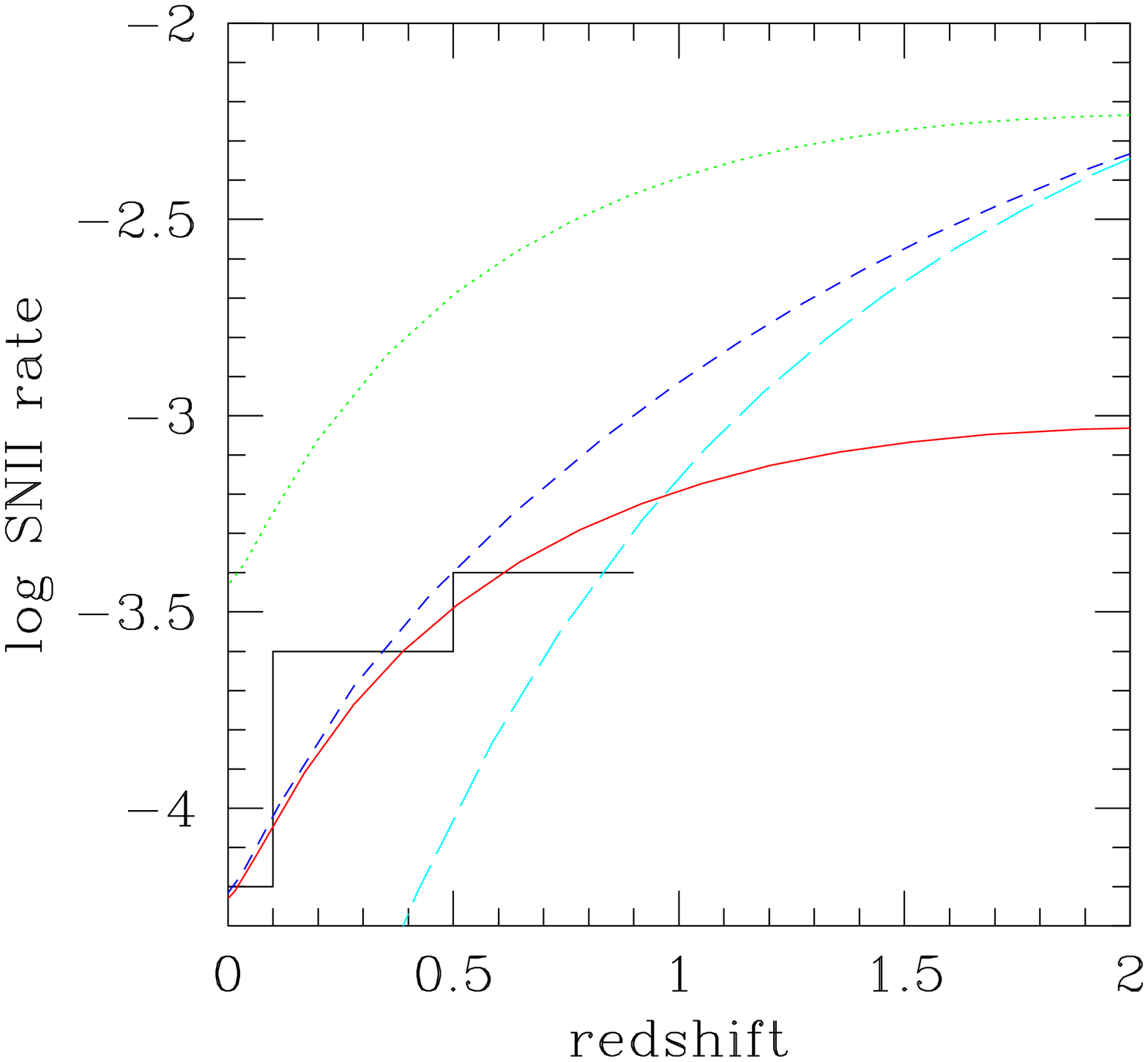}{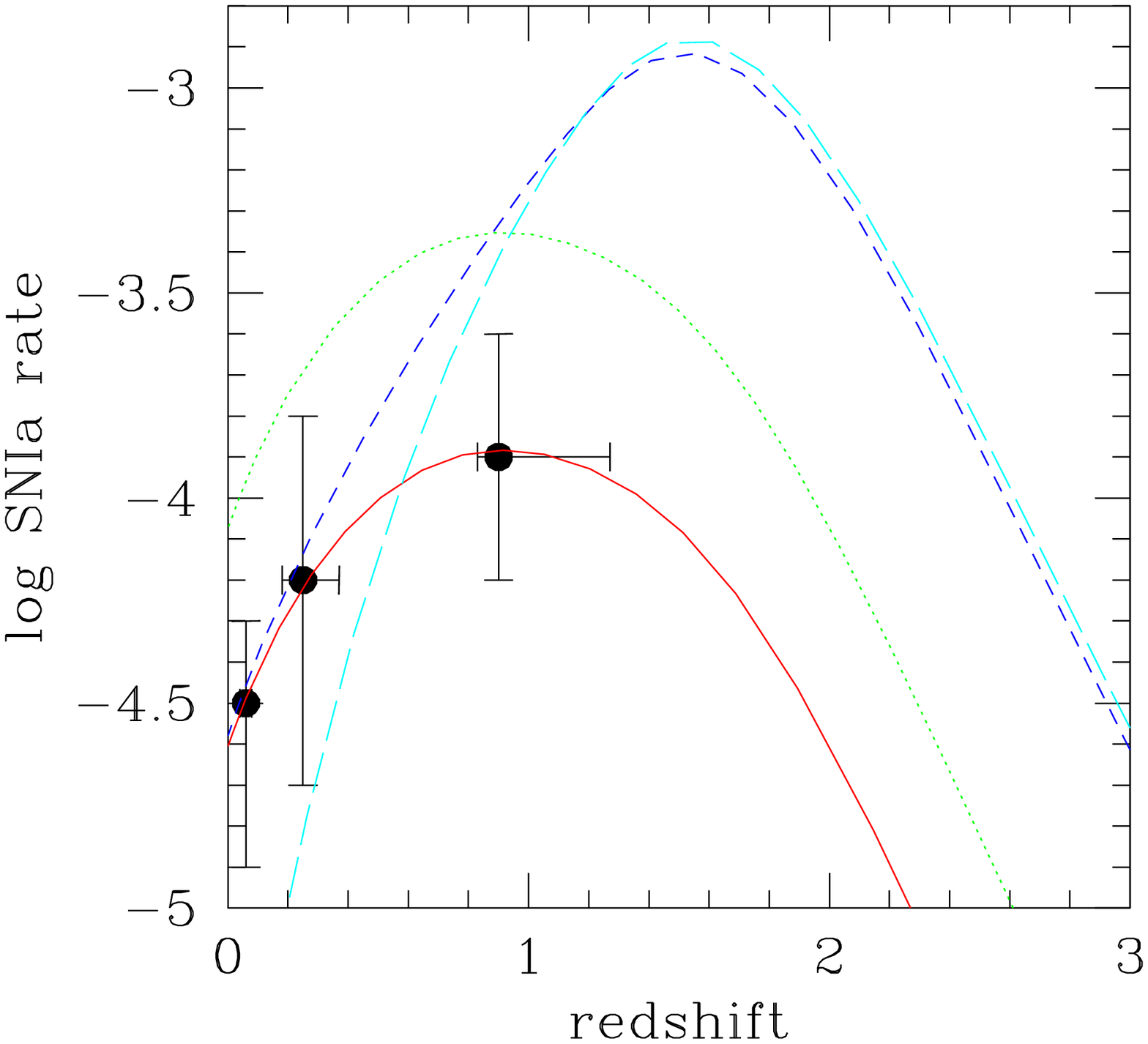}
\caption{left(a): Same as Figure 9 for SNII rate histories, with the
observed field rate histogram reproduced from Figure 3. right(b): Same
as (a) for SNIa rate history. Errorbars are observed cluster SNIa
rates (with $1\sigma$ errors) from Table 7 of \citet{gms}. Supernova
rates are in units of $\delta^{-1}~{\rm Mp}c^{-3}~{\rm yr}^{-1}$. As
in Figure 9, solid curves match the (scaled) observed field
rates. \label{fig10}}
\end{figure}

\clearpage

\begin{figure}
\plottwo{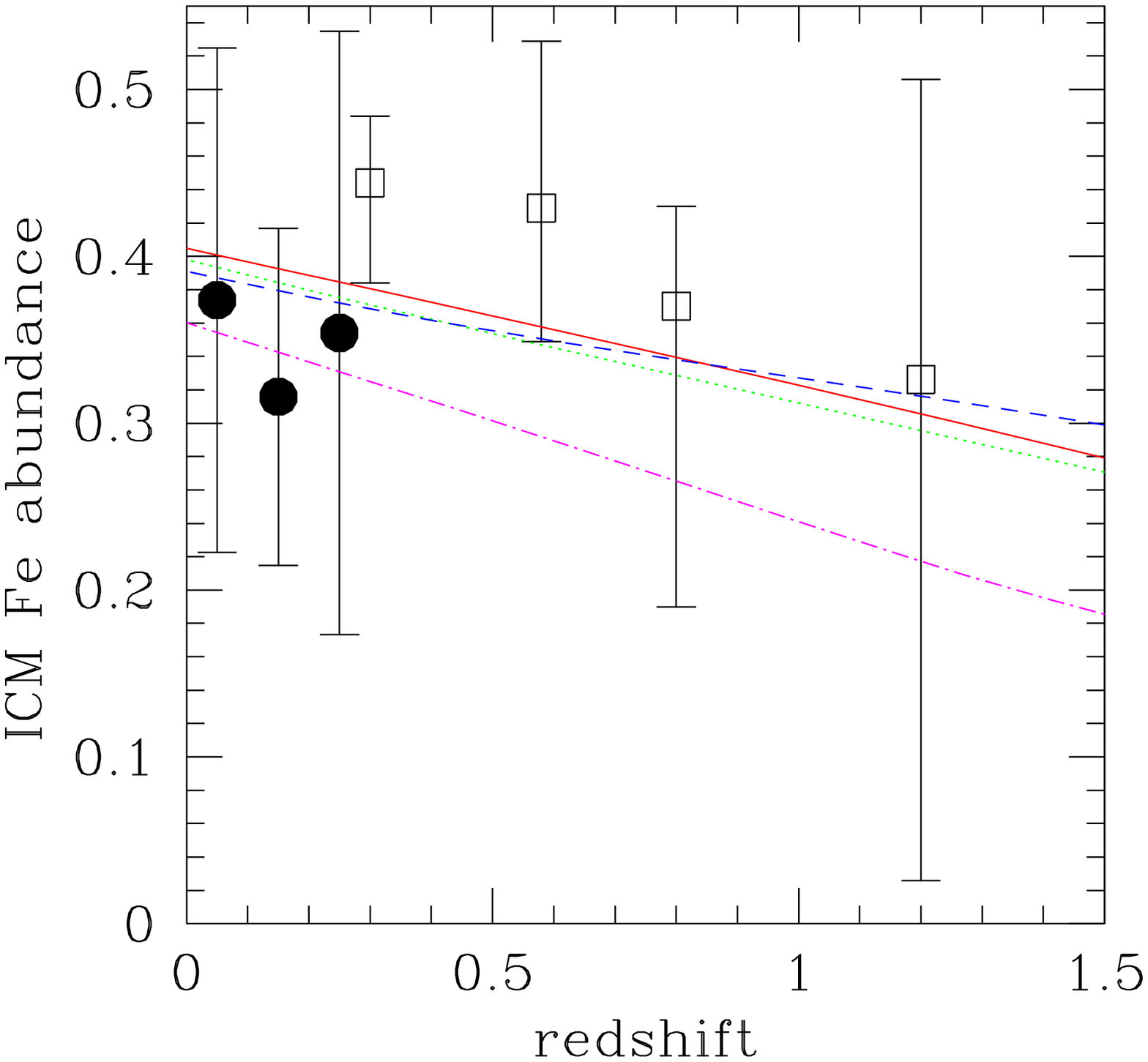}{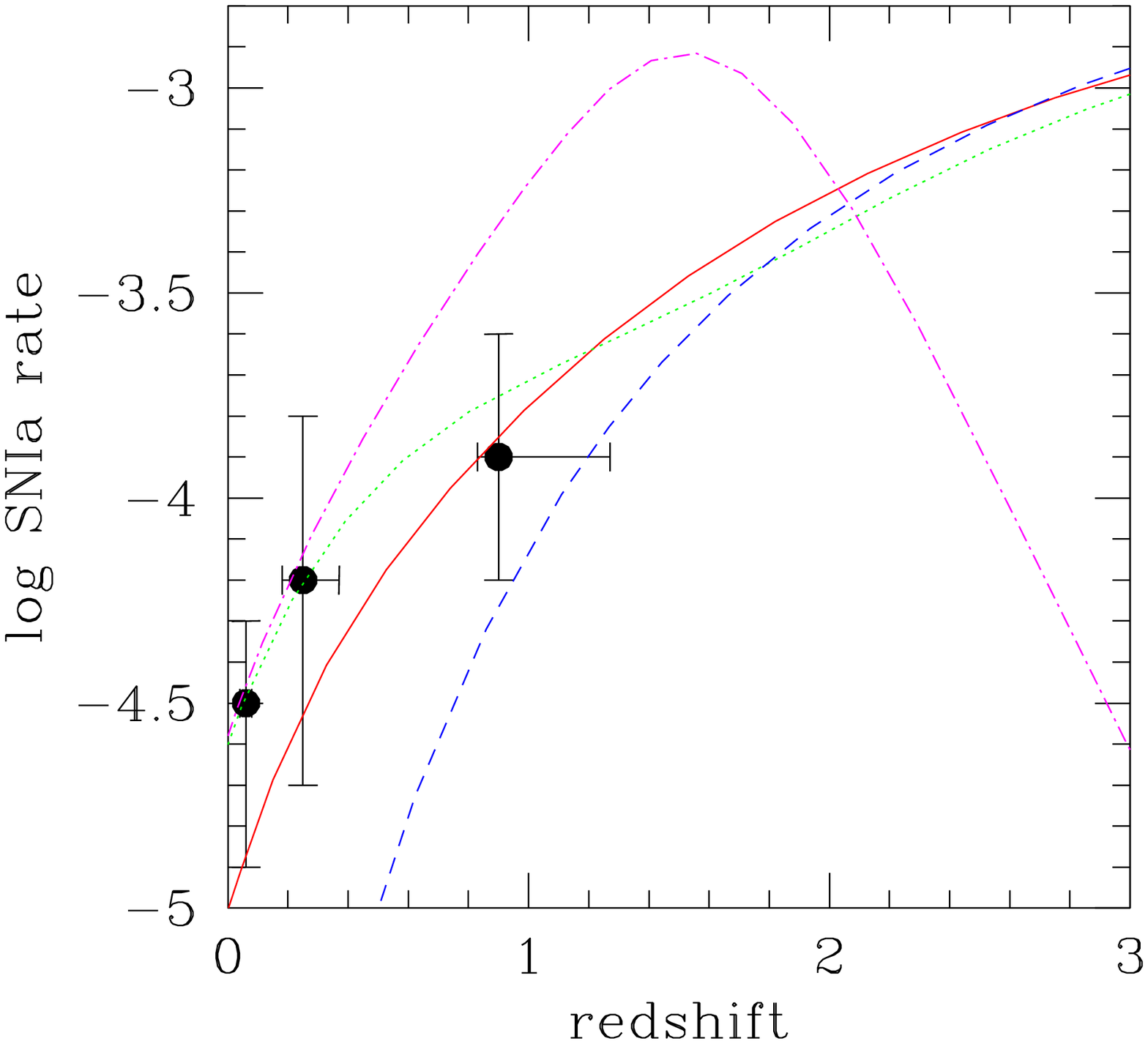}
\caption{left(a): Evolution of ICM Fe abundance for models with
reduced SNIa delay times, 2H1.05WxSt1, 2H1.05WxSt2, 2R1.55WxSt,
(solid, dot, short-dash curves respectively). For comparison, the
evolution for model 2H1.05Wx from Figure 7a is re-plotted (dot-dashed
curve). right(b): Same as (a) for SNIa rate (in $\delta^{-1}~{\rm
Mp}c^{-3}~{\rm yr}^{-1}$) history. \label{fig11}}
\end{figure}

\clearpage

\begin{figure}
\plottwo{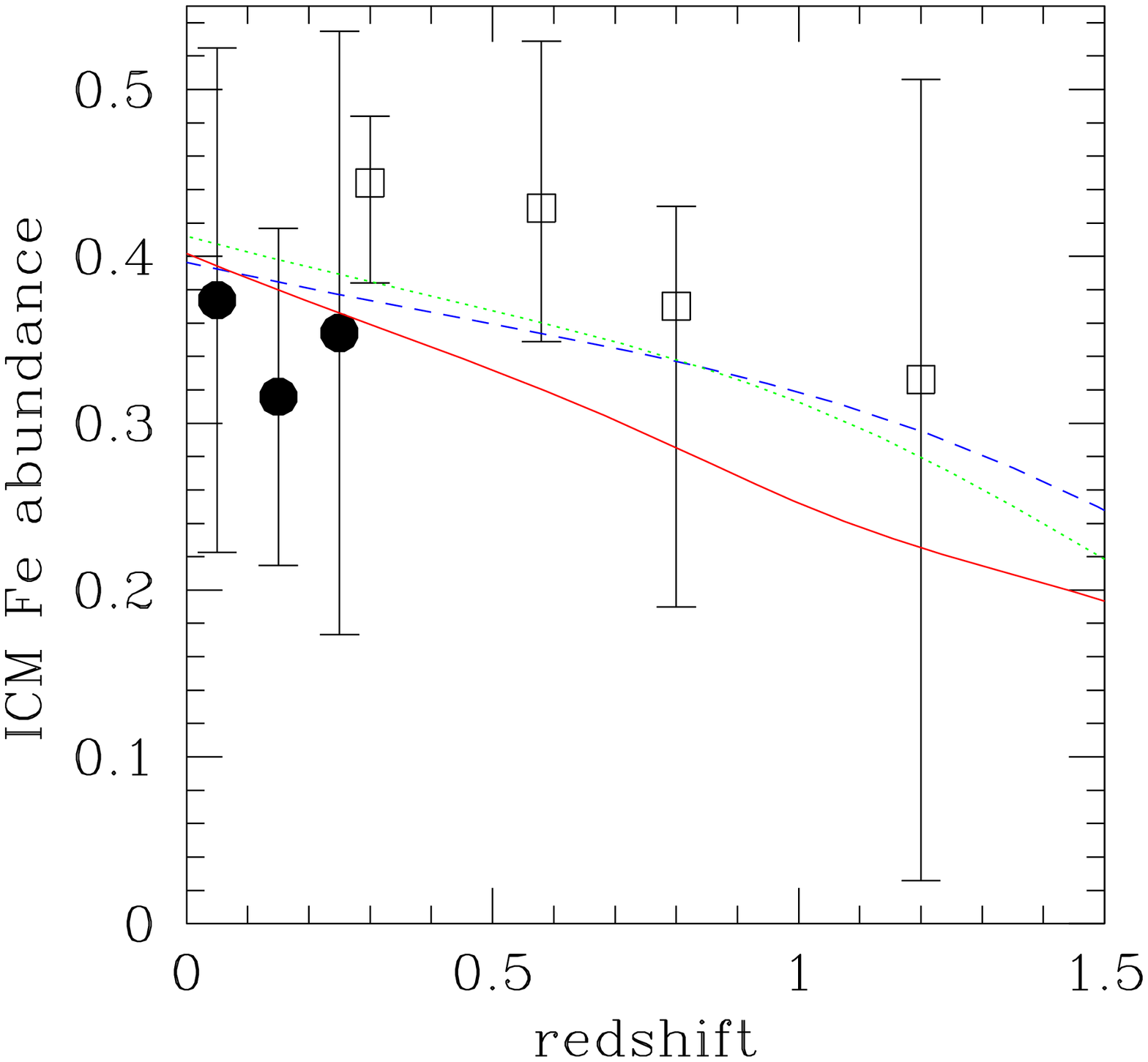}{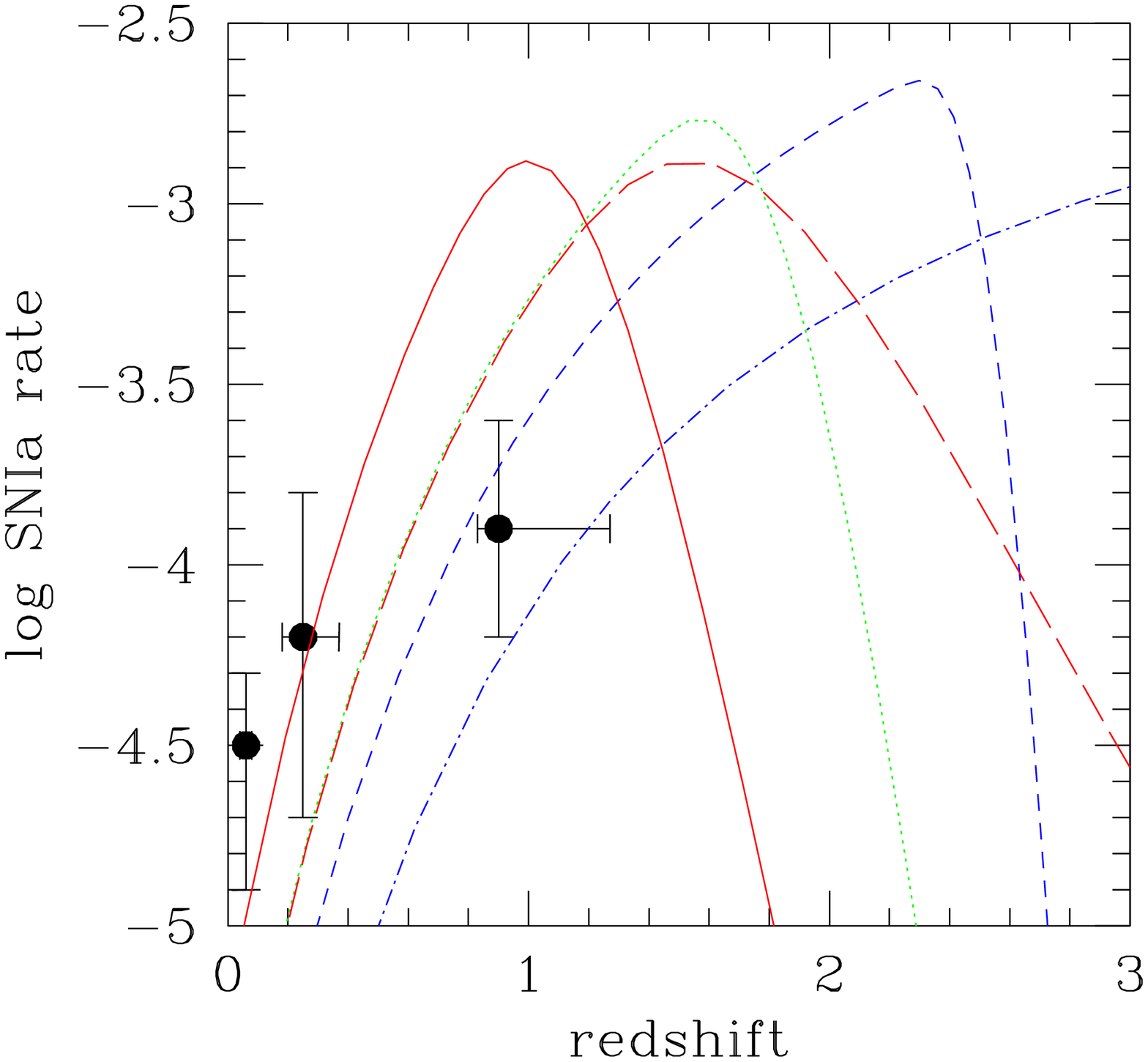}
\caption{left(a): Evolution of ICM Fe abundance for models with rapid
mode star formation initiating at $z=3$, rather than $z=10$.  Shown
are models with mean SNIa delay times of 3 (2R1.55WxSz; solid curve),
1.5 (2R1.55WxStpz; dotted curve), and 0.5 (2R1.55WxStz; short-dashed
curve) Gyr. right(b): Same as (a) for SNIa rate (in $\delta^{-1}~{\rm
Mp}c^{-3}~{\rm yr}^{-1}$). The $z=10$ counterparts for delay times of
3 (2R1.55Wx; long-dashed curve) and 0.5 (2R1.55WxSt; dot-dash curve),
Gyr are reproduced.
\label{fig12}}
\end{figure}

\clearpage

\begin{figure}
\plottwo{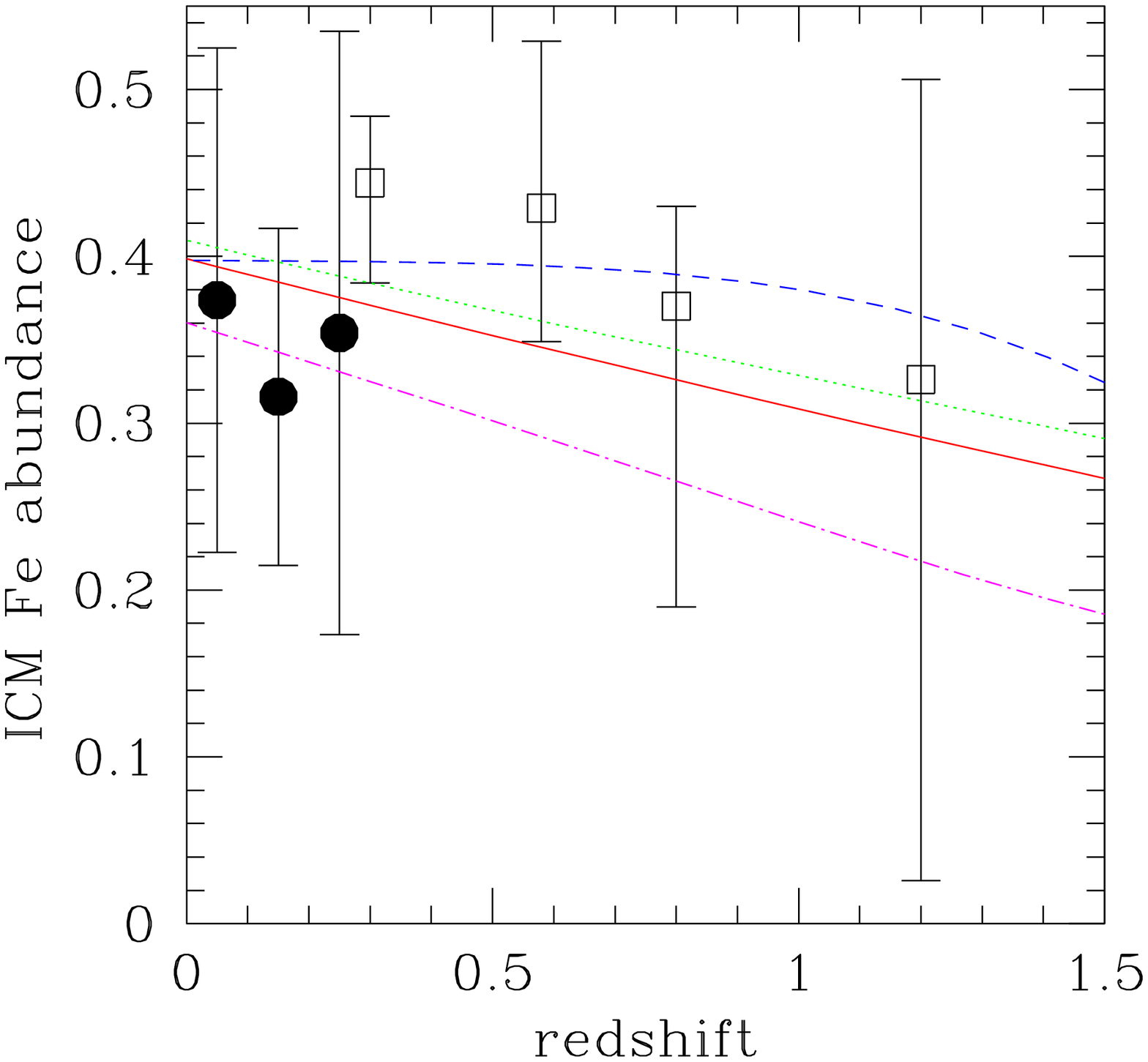}{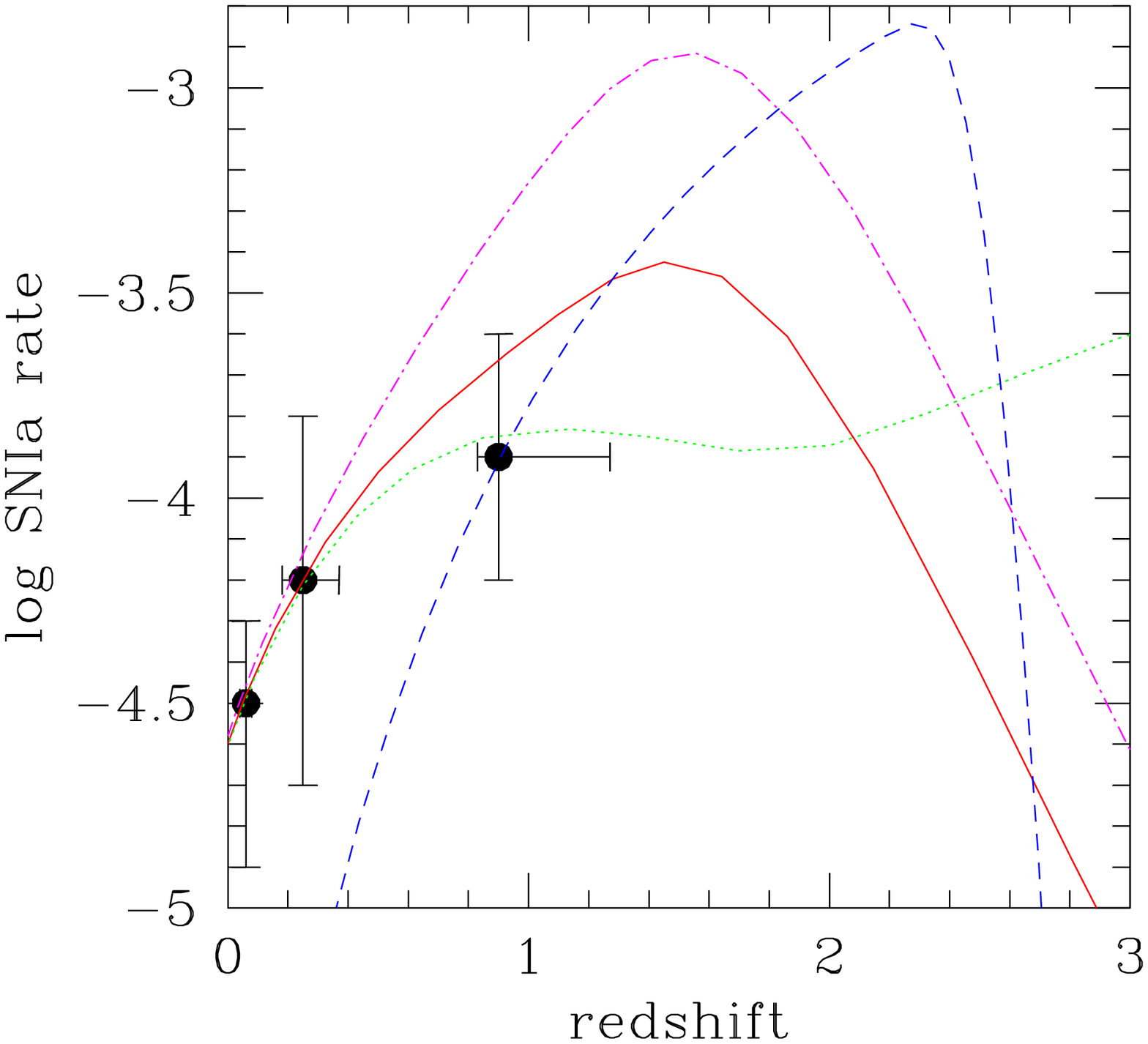}
\caption{left(a): Evolution of ICM Fe abundance with ``naturally''
reduced SNIa normalization and $\langle {y_{\rm SNII}}^{\rm
Fe}\rangle=0.1~{\rm M}_{\odot}$. In hybrid star formation history
models, $\alpha_{3X}=1.05$, $z(t_{\rm form})=10$, and $t_{cX}=3$
(2H1.05WxSn; solid curve) or 0.5 Gyr (2H1.05WxStn; dotted curve); in
the rapid star formation model, $\alpha_{3X}=1.05$, $z(t_{\rm
form})=3,$ and $t_{cX}=0.5$ Gyr (2R1.55Stnz; dashed curve). For
comparison, the evolution for model 2H1.05Wx from Figure 7a is
re-plotted (dot-dashed curve). right(b): Same as (a) for SNIa rate (in
$\delta^{-1}~{\rm Mp}c^{-3}~{\rm yr}^{-1}$) history.
\label{fig13}}
\end{figure}

\clearpage

\begin{figure}
\plotone{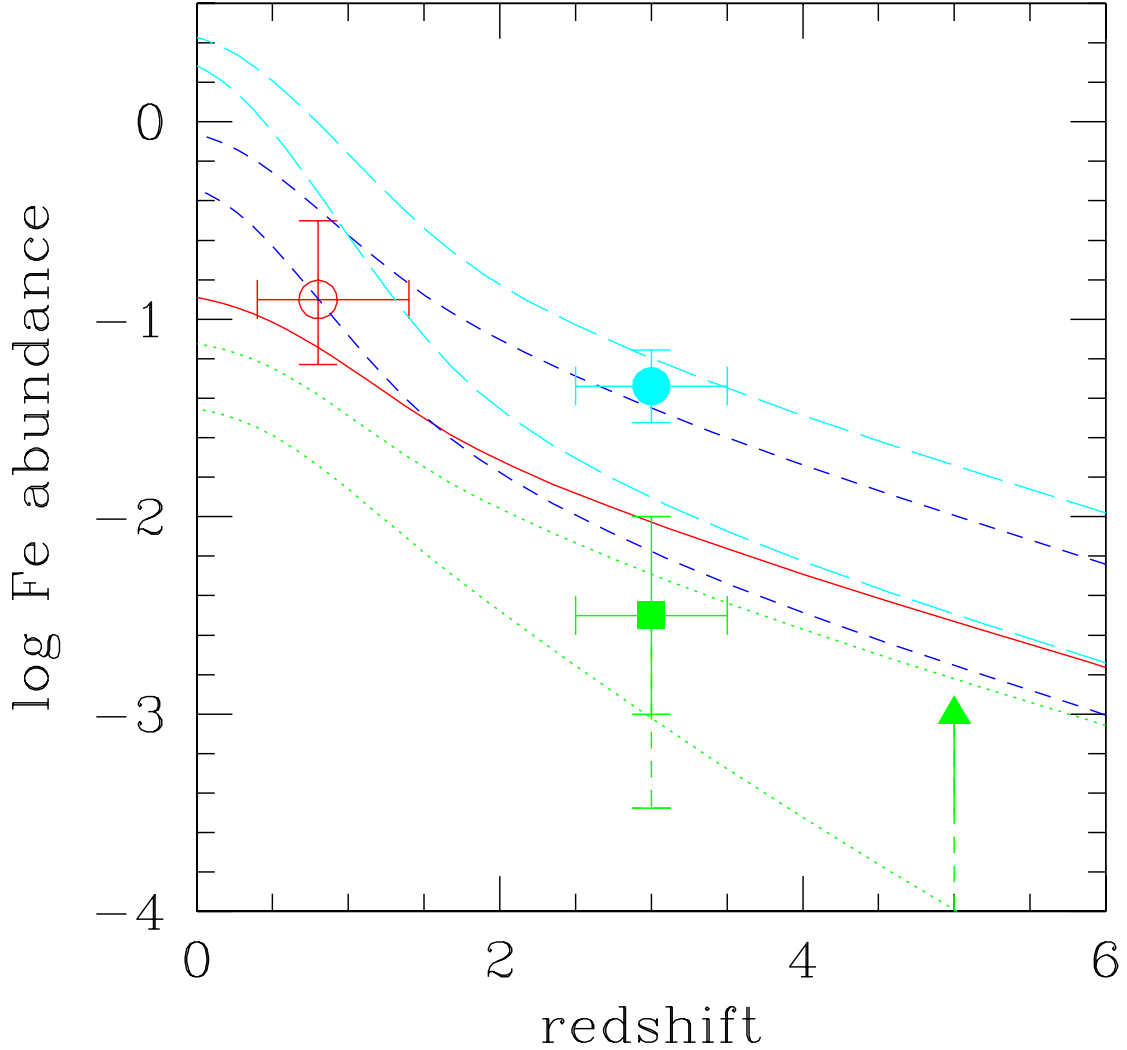}
\caption{Evolution of the average baryon (solid curve), IGM (dotted
curves), stellar (short-dashed curves), and ISM (long-dashed curves)
standard model Fe abundances for the ``maximum wind'' (IGM: upper
curve, stars/ISM: lower curve) and ``intermediate'' cases shown in
Figures 4 and 5. The arrow denotes the $z=5$ Ly$\alpha$ forest lower
limit from \citet{s01}; the solid square with errorbars the range of
$z=2.5-3.5$ Ly$\alpha$ forest measurements from
\citet{s03};~\citet{ssr}, and references therein; the solid circle the
$z=2.5-3.5$ damped Ly$\alpha$ value with errorbars reflecting the
uncertain dust depletion fraction of Fe from \citet{pe04} and
references therein; and, the open circle with errorbars the
$z=0.4-1.5$ damped Ly$\alpha$ systems measurement from
\citet{r05}. The downward broken extensions for the Ly$\alpha$ forest
limits make allowances for the possibility of an $\alpha$/Fe abundance
ratios as high as 3:1.\label{fig14}}
\end{figure}

\clearpage

\begin{figure}
\plottwo{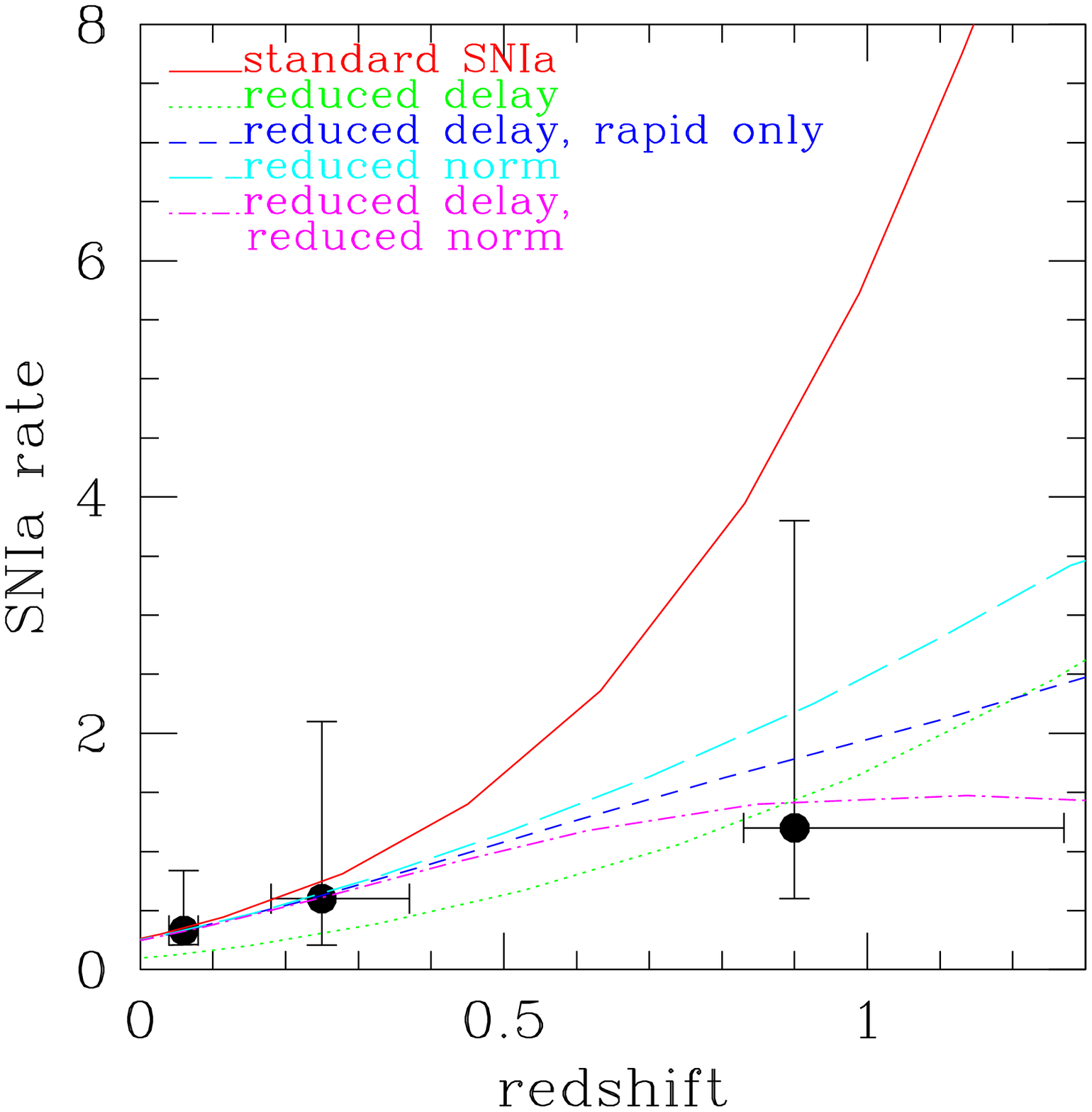}{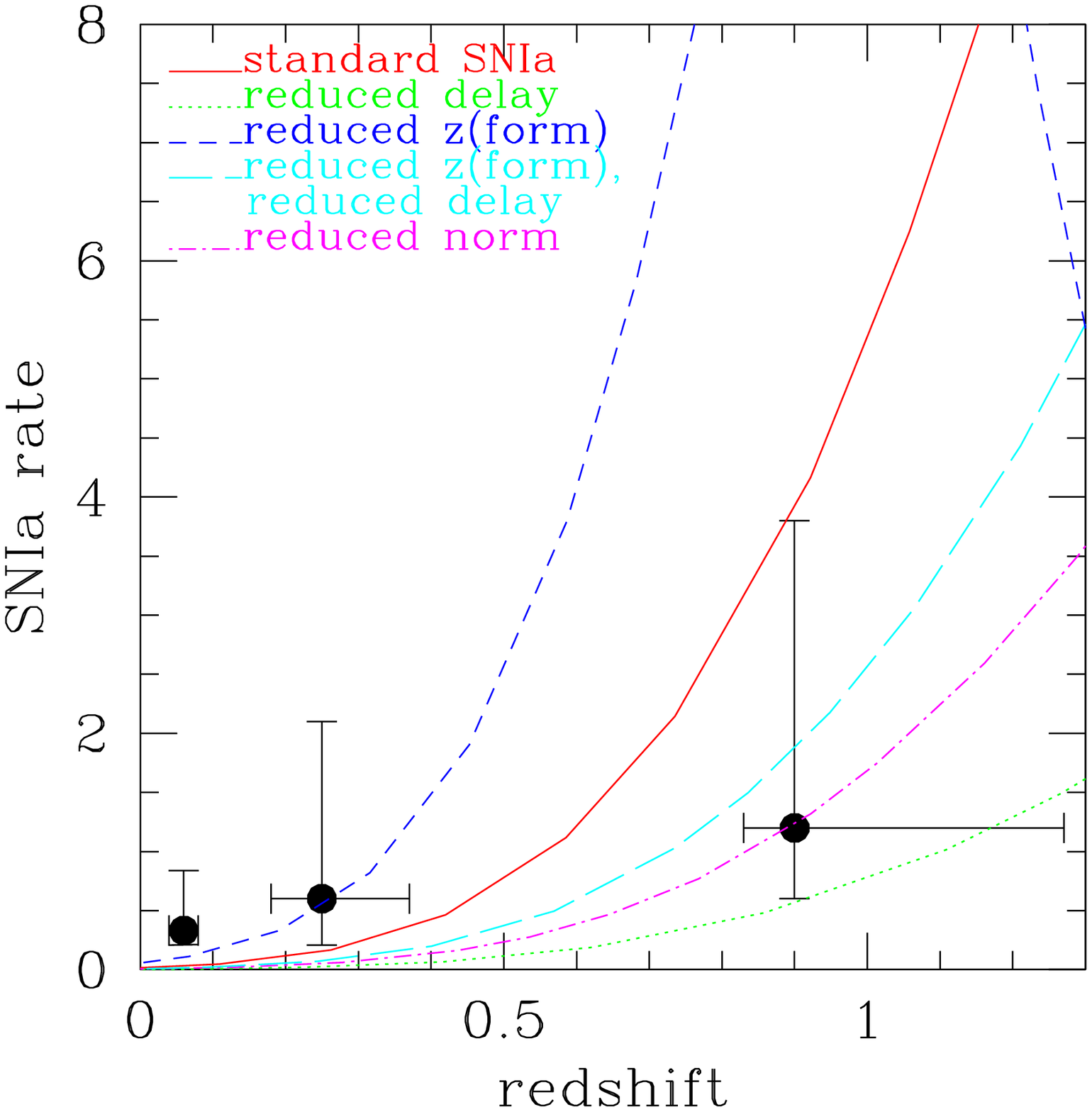}
\caption{left(a): SNIa rate (in $10^{-4}~\delta^{-1}~{\rm
Mp}c^{-3}~{\rm yr}^{-1}$) for Models 2H1.05Wx (solid curve),
2H1.05WxSt1 (dotted curve), 2H1.05WxSt2 (short-dashed curve),
2H1.05WxSn (long-dashed curve), and 2H1.05WxStn (dot-dashed
curve). right(b): Same as (a) for Models 2R1.55Wx (solid curve),
2R1.55WxSt (dotted curve), 2R1.55WxSz (short-dashed curve),
2R1.55WxStz (long-dashed curve), and 2R1.55Stnz (dot-dashed curve).
\label{fig15}}
\end{figure}

\clearpage

\begin{figure}
\plottwo{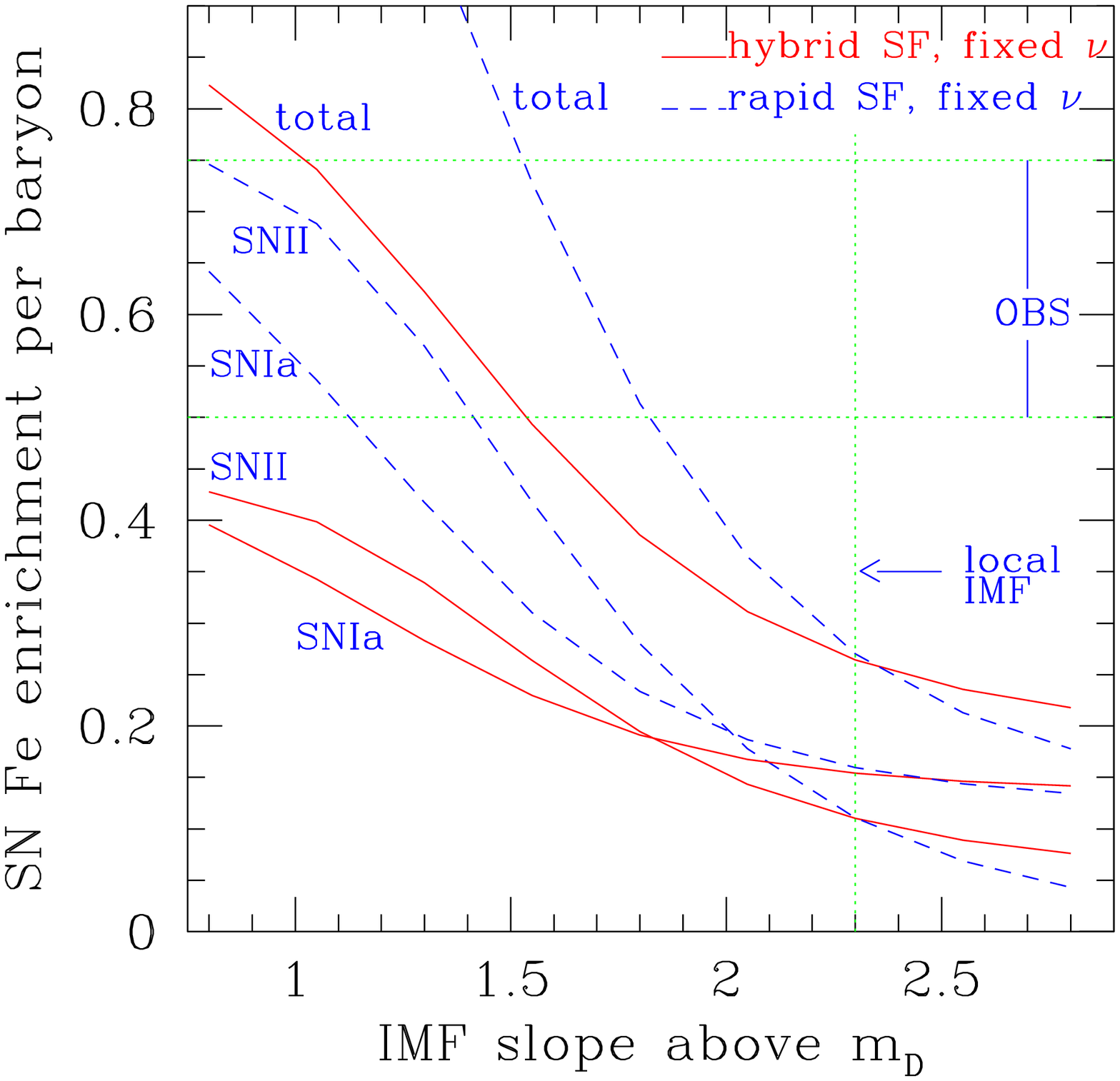}{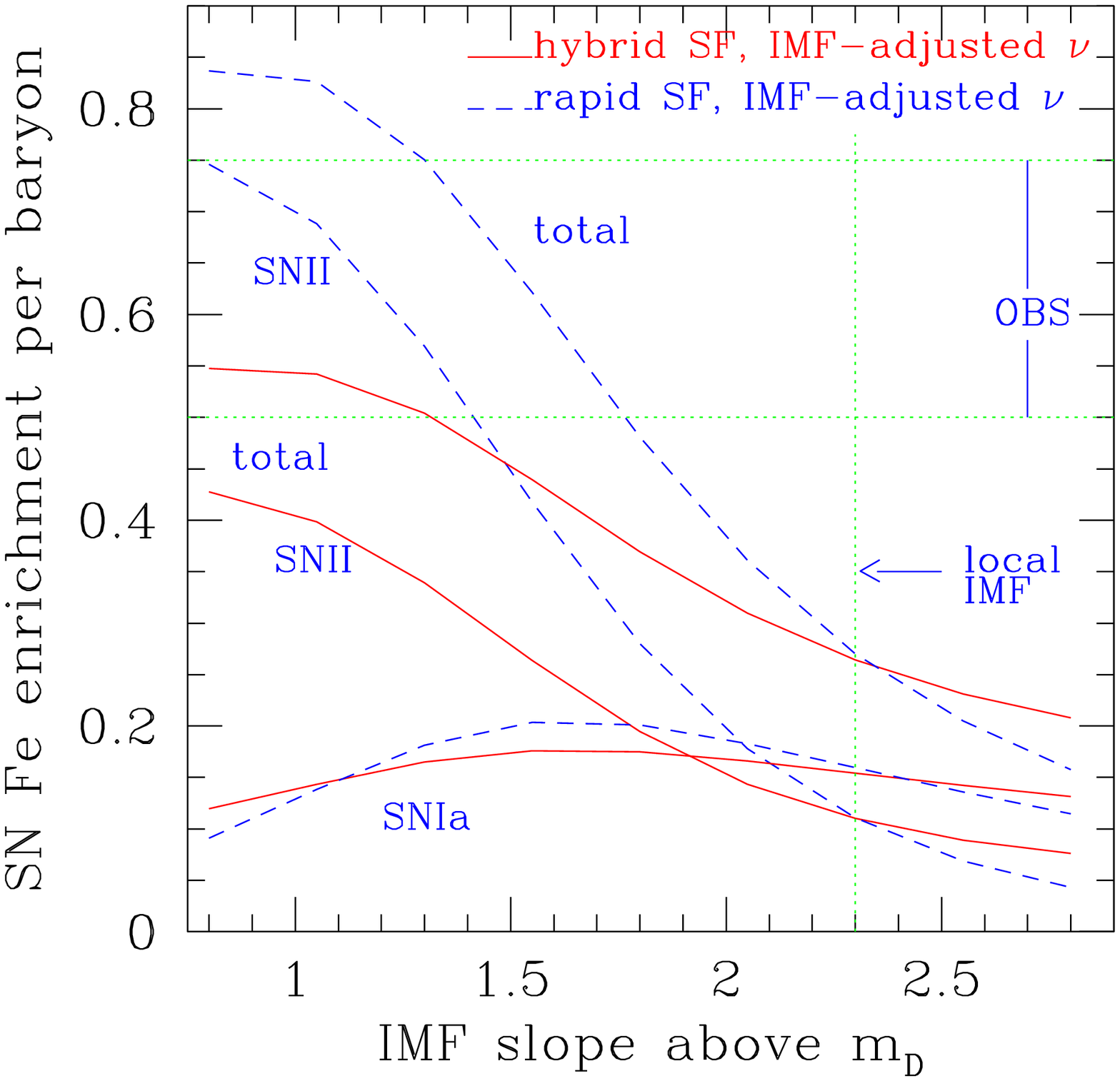}
\caption{left(a): Enrichment per baryon from SNII, SNIa, and their sum
as a function of high mass IMF slope for hybrid and rapid star
formation histories, where the SNIa rate per star formed is fixed at
its field value. The approximate range of observed values is
demarcated by horizontal dotted lines. Curves are calculated assuming
${y_{\rm SNIa}}^{\rm Fe}=0.7~{\rm M}_{\odot}$ and ${y_{\rm SNII}}^{\rm
Fe}=0.07~{\rm M}_{\odot}$. right(b): Same as (a) with the SNIa rate
per star formed scaled according to the IMF.
\label{fig16}}
\end{figure}

\end{document}